\documentclass[epj,nopacs]{svjour}
\usepackage{graphicx,graphics,latexsym,epsfig}
\usepackage{graphicx}
\usepackage{color}
\usepackage{amsmath}
\usepackage{amssymb}
\newcommand{\bc}{\begin{center}}
\newcommand{\ec}{\end{center}}
\newcommand{\be}{\begin{eqnarray}}
\newcommand{\ee}{\end{eqnarray}}

\newcommand\scatt{\mathrm{scatt}}

\newcommand{\nn}{\nonumber}
\begin{document}
\title{\bf Further study of $D^+_s$ decays into
$\pi^- \pi^+ \pi^+$}
\author
{ E. Klempt$\,^1$ \and M. Matveev$\,^{1,2}$ \and A.V.
Sarantsev$\,^{1,2}$ }

\institute
{Helmholtz--Institut f\"ur Strahlen-- und Kernphysik,
Universit\"at Bonn, Germany
\and
Petersburg Nuclear Physics Institute, Gatchina, Russia
}
\date{Received: July 23, 2007 / Revised version: February 29, 2008}

\titlerunning{Further study of $D^+_s$ decays into
$\pi^- \pi^+ \pi^+$}

\abstract{ A Dalitz plot analysis of the OZI rule violating decay
$D^+_s$ into $\pi^- \pi^+ \pi^+$ is presented using different
partial wave approaches. Scalar and vector waves are described by
$K$-matrices; their production is parameterized in a $P$-vector
approach. Alternatively, Breit-Wigner amplitudes and Flatt\'e
parametrization are used. Special emphasis is devoted to scalar
mesons. The $f_0(980)$ resonance provides the most significant
contribution. Adding $f_0(1500)$ to the scalar wave leads to an
acceptable fit while introduction of $f_0(1370)$ and/or $f_0(1710)$
does not lead to significant improvements. A scan of the scalar wave
optimizes for $M=1452\pm 22$\,MeV/c$^2$. When $f_0(1710)$ is added,
the mass uncertainty increases, and the fit yields $M=1470\pm
60$\,MeV/c$^2$ which is fully compatible with the nominal
$f_0(1500)$ mass. The scalar wave seems to exhibit a phase motion of
270$^{\circ}$ units in the mass range from 1200 to
1650\,MeV/c$^2$.\\
\vskip 3mm\vspace{-3mm} PACS:{
       {11.80.Et}{ Partial-wave analysis} \and
       {13.20.Fc}{Decays of charmed mesons} \and
       {14.40.Cs}{Other mesons with S=C=0, mass $< 2.5$\,GeV }
}} \mail{klempt@hiskp.uni-bonn.de}

\authorrunning{E. Klempt,  M. Matveev, and A.V. Sarantsev}
\titlerunning{Further study of $D^+_s$ decays into
$\pi^- \pi^+ \pi^+$}
\maketitle

\section{Introduction}

Decays of charmed mesons provide an efficient tool to study
meson-meson interactions at low energies.  Outstanding examples are
the analyses of the $D^+\to \pi^-\pi^+\pi^+$ Dalitz plot which
helped to establish the $\sigma(500)$ \cite{Aitala:2000xu}, and the
two reactions $D^0\to K^0_S\pi^+\pi^-$ \cite{Muramatsu:2002jp} and
$D^+\to K^-\pi^+\pi^+$ \cite{Aitala:2002kr,Aitala:2005yh} which
revealed the existence of the $\kappa(700)$ meson. The reaction
$D_s^+ \to \pi^- \pi^+ \pi^+$ was proven to provide access to mesons
in which the primarily formed $s\bar s$ state converts into a
$\pi^-\pi^+$ pair
\cite{Frabetti:1997sx,Aitala:2000xt,Link:2003gb,Malvezzi:2004tq} in
a OZI rule violating transition. In $D$ and $D_s$ decays into three
pseudoscalar mesons, a large fraction of the cross section is
assigned to a pseudoscalar meson recoiling against a scalar meson;
this fact makes $D$ and $D_s$ decays very well suited for
investigations of the spectrum of scalar mesons and their flavor
wave function. A survey of data sets and of Dalitz plot analyses of
$D$ and $D_s$ decays \cite{Asner:2003gh} can be found in the Review
of Particle Properties \cite{PDG}.

In spite of the large potential for illuminating contributions to
our understanding of scalar mesons up to a mass of $\sim
1700$\,MeV/c$^2$, the situation concerning the states $f_0(1370)$
and $f_0(1500)$ and their contributions to the reaction $D_s^+ \to
\pi^- \pi^+\pi^+$ is still controversial. The evidence for the
existence of $f_0(1370)$ has often been questioned
\cite{Estabrooks:1978de,Au:1986vs,Minkowski:1998mf,Klempt:2007cp};
if it exists, the flavor decomposition of the scalar states remains
unclear as evidenced by the numerous mixing schemes in which a
scalar glueball is supposed to intrude into the spectrum of scalar
$q\bar q$ meson, to mix with them thus creating the observed pattern
of scalar resonances. Instead of a `narrow' $f_0(1370)$, a wide
scalar background has been proposed which was called $f_0(1000)$ by
Au, Morgan and Pennington \cite{Au:1986vs}, and `red dragon' by
Minkowski and Ochs \cite{Minkowski:1998mf}. A survey of different
mixing schemes and a critical discussion of their foundations are
reported in a recent review \cite{Klempt:2007cp}.

All analyses of $D^+_s$ decays into $\pi^- \pi^+ \pi^+$ agree on
basic features of the data even though different partial wave
analysis techniques were applied which led -- in important details
-- to rather different conclusions
\cite{Frabetti:1997sx,Aitala:2000xt,Link:2003gb,Malvezzi:2004tq}. In
all analyses the Dalitz plot is shown to be dominated by $f_0(980)$.
There is possibly a small contribution from $\pi^+\rho^0$;
significant contributions stem from $\pi^+\rho(1450)$ and
$\pi^+f_2(1270)$. In the scalar isoscalar partial wave, there is a
sizable contribution which leads to a peak at a mass of about 1450
MeV/c$^2$. The origin of this enhancement is however controversial.
E791 fitted the $D_s^+\to \pi^+\pi^+\pi^-$ data with a scalar
resonance for which $m_0 = $$1434 \pm 18 \pm 9$\,MeV/c$^2$ and
$\Gamma_0 = 173 \pm 32 \pm 6$\,MeV/c$^2$ were found. They had
observed $f_0(1370)$ production in an analysis of the $D\to 3\pi$
Dalitz plot and identified the scalar intensity with $f_0(1370)$.
The $f_0(1370)$ is not supposed to have a large $s\bar s$ component,
but its production could be assigned to the annihilation diagram
(see \ref{fig:d-decay}c). The Focus collaboration reported $m_0 =
$$1475 \pm 10$\,MeV/c$^2$ and $\Gamma_0 = 112 \pm 24$\,MeV/c$^2$.
These parameters are nearly compatible with $f_0(1500)$. Thus, the
important issue to which extend the two states $f_0(1370)$ and
$f_0(1500)$ contribute remained unsettled.

In this paper we report on a further study of the $D_s^+ \to \pi^-
\pi^+ \pi^+$ Dalitz plot. The main point of the analysis is an
exploration of different methods and of the impact the analysis
technique has on the final result.

The paper is organized as follows. In section  \ref{Decays} we remind
the reader of some basic features of weak decays of charmed strange
mesons. Subsequently, a short survey is given of the apparatus (section
\ref{Exp}), of the data and of data selection. Section \ref{PWA}
contains a description of the amplitudes used to fit the data, and the
fit results. The paper ends with a discussion of the results and a
short summary.

\section{\label{Decays}Decays of charmed mesons}
Before starting a partial wave analysis of  $D_s^+\to \pi^-\pi^+\pi^+$
decays (inclusion of the charge conjugate reaction is understood), it
may be useful to remind of some properties of weak decays of charmed
mesons relevant to light-quark spectroscopy.

\begin{figure}[ph]\vspace{-2mm} \begin{center}
\begin{tabular}{cccc}
\includegraphics[width=0.14\textwidth,height=17mm]{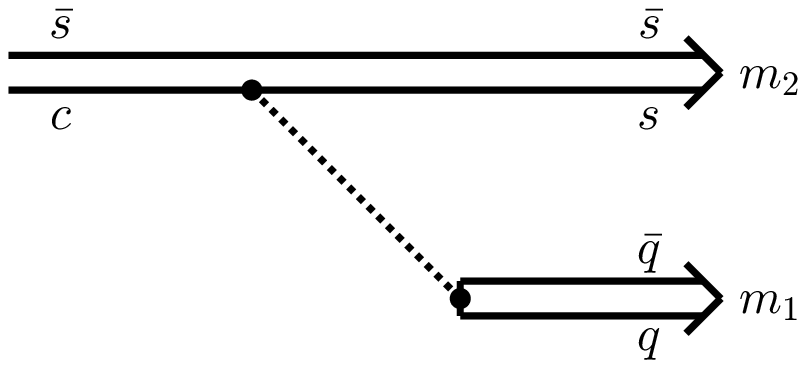}&
\includegraphics[width=0.14\textwidth,height=17mm]{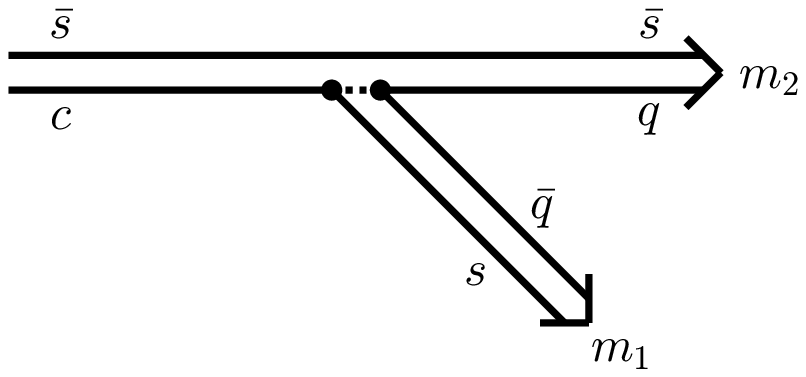}&
\includegraphics[width=0.14\textwidth,height=17mm]{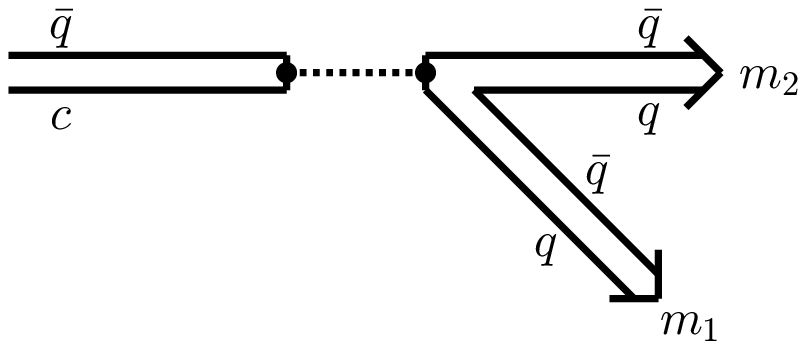}
\end{tabular}

(a)\hspace{20mm}(b)\hspace{20mm}(c) \vspace{-2mm}\end{center}
\caption{\label{fig:d-decay} Decays of charmed meson: a) Leading
mechanism. The $q\bar q$ pair can be $u\bar d$ or $u\bar s$. b)
Color suppressed diagram. The $q\bar q$ pair is likely $u\bar d$. c)
Annihilation diagram. Here, $q$ can be $u,d$ or $s$ quarks. }
\end{figure}

Fig. \ref{fig:d-decay} depicts Feynman diagrams for the decay of
char\-med mesons. In Cabibbo favored decays of $D_s^+$ mesons (Fig.
\ref{fig:d-decay}a) the primary $c$ quark converts into an $s$ quark
under emission of a $W^+$ while the $\bar s$ quark acts as a
spectator
 particle. The most likely decay of the $W^+$ produces
$\pi^+$. Conversion of a $c$ quark into a $d$ quark or $W^+$ decay
into $K^+$ are Cabibbo suppressed. Hence most likely, a meson with
hidden strangeness emerges, recoiling against a $\pi^+$. Evidence
for this diagram can be found in a comparison of $D_s^+\to\pi^+\phi$
and $D_s^+\to\pi^+\omega$. The former reaction occurs with
$(4.4\pm0.6)\cdot10^{-2}$ frequency, the latter reaction with
$(3.4\pm1.2)\cdot10^{-3}$: the reaction $D_s^+\to\pi^+\omega$ is
suppressed. The suppression is not as large as expected from
$\phi$--$\omega$ mixing (for a deviation $\delta_V\sim 3.5^{\circ}$
from the ideal mixing angle, a suppression by
$arctg^{-2}\delta_V\sim1/250$), suggesting other mechanisms to
contribute to $D_s^+\to\pi^+\omega$. These branching fractions and a
few further ones are collected in Table~\ref{dsdecay}.

Fig. \ref{fig:d-decay}b shows a Cabibbo allowed but color suppressed
diagram. The two quarks of the $u\bar d$ pair need to match in color
with the $s$ and the $\bar s$ quark, leading to an expected $1/N_c$
reduction of the probability. However, $D_s^+$ mesons are observed
to decay into $K^+\bar K^{*0}$ and $K^{*+}\bar K^{0}$ with a mean
rate of $(7.9\pm 1.4)$\%, to be compared with the
$(4.4\pm0.6)\cdot10^{-2}$ frequency with which $D_s^+\to\pi^+\phi$
is produced. Thus additional processes contributing to production of
open strangeness are required.

The annihilation diagram of Fig. \ref{fig:d-decay}c contributes to
$D_s^+$ decays with an a priori unknown fraction. The intermediate
$W^+$ may convert into the 3$\pi$ final state via $\rho\pi^+$ or
$f_2(1270)\pi^+$ but also into a scalar and a pseudoscalar meson.

Evidence for annihilation can be deduced from leptonic decays modes
\cite{PDG}. $D_s^+$ mesons have a large probability to decay into
$\tau^+\nu_{\tau}$; decays into $\mu^+\nu_{\mu}$ or $e^+\nu_{e}$ are
suppressed due to helicity conservation. The $u,d$ (constituent)
quark masses suggest hadronic branching ratios to contribute several
\% when the number of colors is included. $D_s^+\to\pi^+\omega$
production is thus likely due to the annihilation diagram. The
strange mass is of course larger; annihilation could be the reason
for the unexpectedly high $K^+\bar K^{*0}$ and $K^{*+}\bar K^{0}$
yields. Cabibbo and color suppressed decays could also contribute to
$\pi^+\omega$ production by rescattering processes in the final
state; these processes are however expected to make at most small
contributions.

Large fractional yields in purely pionic final states are observed via
production of $\eta$ and $\eta'$ in the intermediate state. It is clear
why these yields are comparatively large: due  to their
$s\bar s$ components, $\eta$ and $\eta'$ couple strongly to the initial
$s\bar s$ system. Their $n\bar n$ component leads to pionic final
states. The large yields of scalar mesons in $D_s^+$ decays points to
the presence of both, $n\bar n$ and $s\bar s$ components in the wave
functions of scalar mesons.

\begin{table}[pt]
\caption{\label{dsdecay}Selected
$D_s^+$ decay modes \cite{PDG}}
\renewcommand{\arraystretch}{1.3}
\begin{scriptsize}
\begin{tabular}{lrclr}
\hline\hline
\hspace{-1mm}$D_s^+\to\mu^+\nu_{\mu}$&
\hspace{-3mm}$(6.1\pm1.9)\cdot10^{-3}$&\hspace{-1mm}&
\hspace{-3mm}$D_s^+\to\tau^+\nu_{\tau}$&
\hspace{-3mm}$(6.4\pm1.5)\cdot10^{-2}$\hspace{-6mm}\\
\hspace{-1mm}$D_s^+\to\pi^+\phi$&
\hspace{-3mm}$(4.4\pm0.6)\cdot10^{-2}$&\hspace{-1mm}&
\hspace{-3mm}$D_s^+\to\pi^+\omega$&
\hspace{-3mm}$(3.4\pm1.2)\cdot10^{-3}$\hspace{-6mm}\\
\hspace{-1mm}$D_s^+\to\pi^+\eta$&
\hspace{-3mm}$(2.11\pm0.35)\cdot10^{-2}$&\hspace{-1mm}&
\hspace{-3mm}$D_s^+\to\pi^+\eta'$&
\hspace{-3mm}$(4.7\pm0.7)\cdot10^{-2}$\hspace{-6mm}\\
\hspace{-1mm}$D_s^+\to\rho^+\eta$&
\hspace{-3mm}$(13.1\pm2.6)\cdot10^{-2}$&\hspace{-1mm}&
\hspace{-3mm}$D_s^+\to\rho^+\eta'$&
\hspace{-3mm}$(12.2\pm2.4)\cdot10^{-2}$\hspace{-6mm}\\
\hline\hline
\end{tabular}
\end{scriptsize}
\renewcommand{\arraystretch}{1.0}
\end{table}

\section{\label{Exp}Detector and data analysis}
\subsection{The \boldmath$E791$\unboldmath\  experiment}
The data analyzed here have been collected at Fermilab in a 500
GeV/$c$ $\pi^- $ beam impinging on platinum and carbon targets. The
pion beam was tracked in proportional wire chambers (PWC's) and
silicon microstrip detectors (SMD's) in front of the targets;
particles emerging from a hadronic reaction were detected in a
spectrometer which consisted of further PWC's and SMD's, two
magnets, 35 drift chamber (DC) planes, two gas \v Cerenkov counters,
an electromagnetic calorimeter, a hadronic calorimeter and a muon
detector composed of an iron shield and two planes of scintillation
counters. A full description of the detector, of data reconstruction
and of data analysis can be found in
\cite{Aitala:2000xt,Aitala:1998kh}.

\begin{figure}[pt]
\begin{center}
\begin{tabular}{cc}
\includegraphics[width=0.24\textwidth]{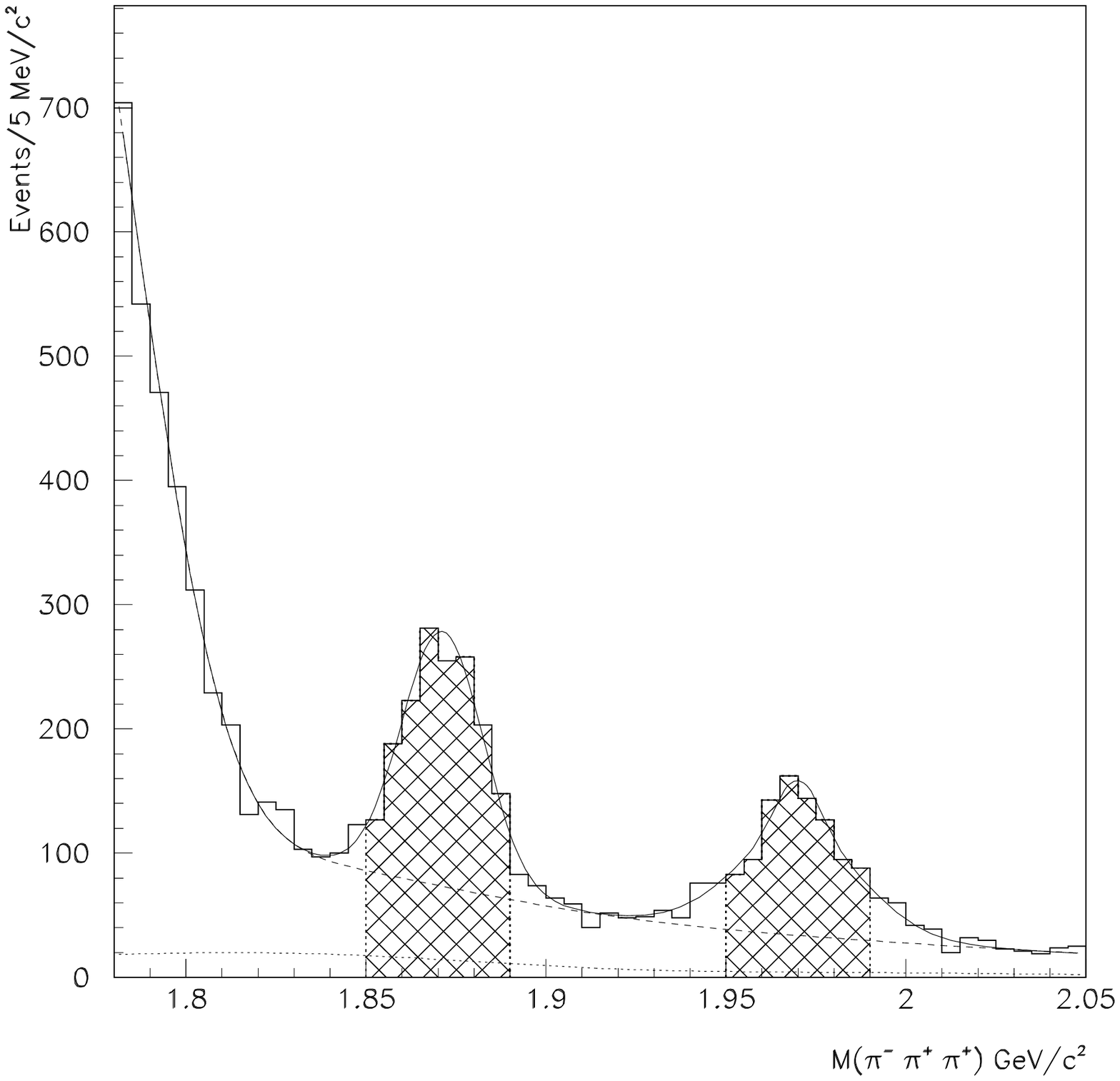}&
\hspace{-5mm}\includegraphics[width=0.24\textwidth]{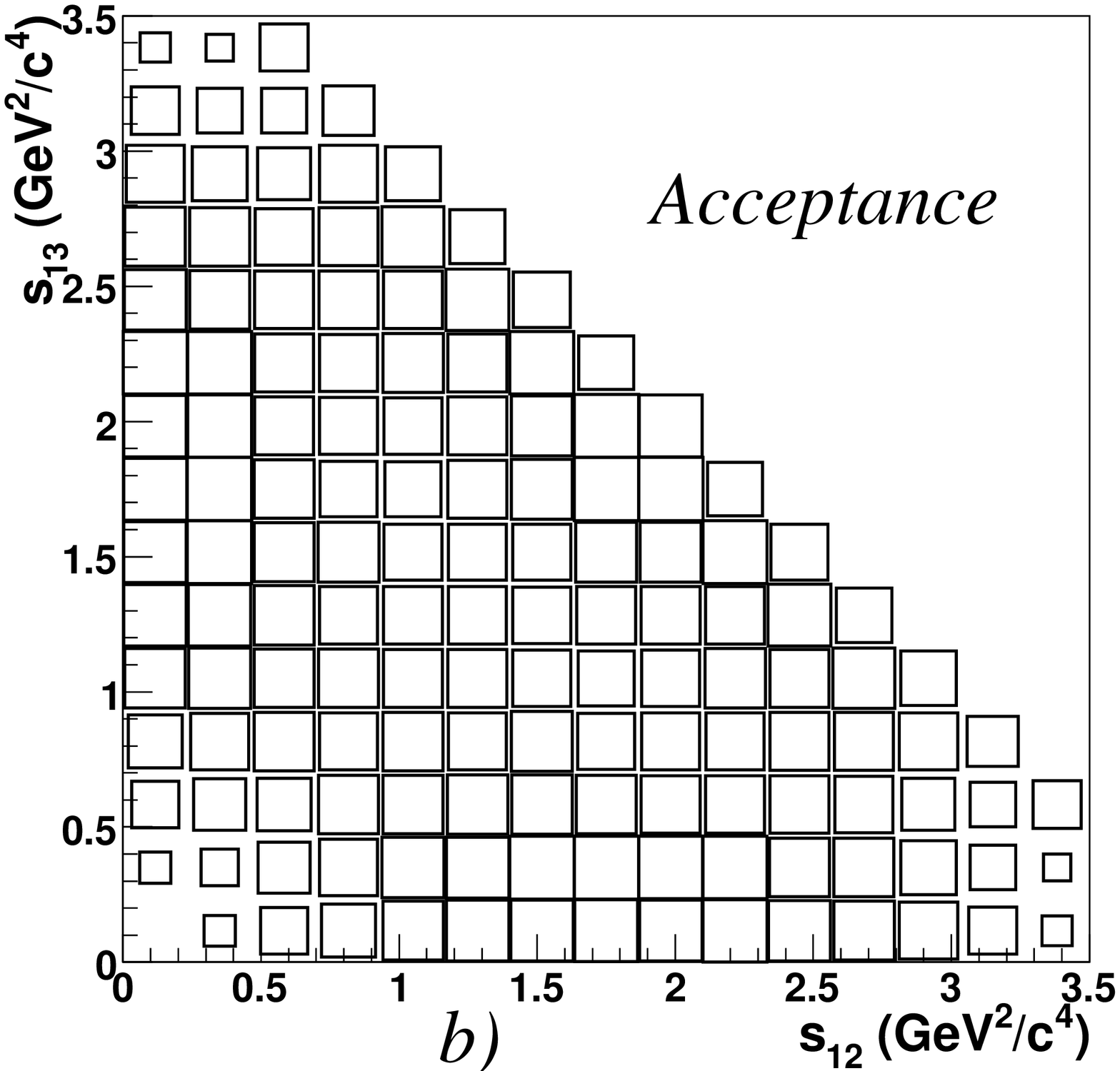}
\end{tabular}\vspace{-4mm}\\
\hspace{15mm}$a)$\hspace{62mm} \vspace{1mm}
\end{center}
\caption{\label{massdistr} a: The $\pi^- \pi^+ \pi^+$ invariant mass
spectrum. The dotted line represents the sum of the expected
contributions from $D^0\to K^-\pi^+$ and $D_s^+ \to \eta^{\prime}
\pi^+$ decays; the dashed line is the total background. Events in
the hatched areas at the $D_s$ mass are used in this analysis. The
hatched area at the $D^+$ mass was used to establish the
$\sigma(500)$ and its properties. b: Detector acceptance in the mass
range from 1.97 to 1.99\,GeV/c$^2$.} \end{figure}

In a first analysis stage, events were required to have a
reconstructed primary production vertex whose location coincided
with one of the target foils. Furthermore, events had to have a well
separated secondary decay vertex (more than 4$\sigma_{\rm
vertex~recon~error}$ in the longitudinal separation), and 3
reconstructed tracks with a total charge $+1$. Particle
identification was not required, tracks were assumed to originate
from charged pions. The mass distribution of events due to
three-pion systems surviving these cuts is shown in Fig.
\ref{massdistr} for $\pi^-\pi^+\pi^+$ invariant masses falling into
the 1.7 to 2.1\,GeV/c$^2$ mass bin. The data can be described by two
Gaussians at 1.87 and 1.97\,GeV/c$^2$ with resolution
$\sigma_{1.87}=11.8$ and $\sigma_{1.97}=13.5$\,MeV/c$^2$,
respectively. For the background function, an exponential form was
chosen. There are $1172\pm 61$ events due to $D\to \pi^-\pi^+\pi^+$
events and  $848\pm 44$ events due $D_s^+ \to \pi^-\pi^+ \pi^+$.

After a cut in the $\pi^-\pi^+\pi^+$ invariant mass between 1.95 and
1.99 GeV/c$^2$ as indicated in Fig. \ref{massdistr}a, $937$ events
are retained for further analysis. The integrated signal to
background ratio was estimated to $\sim 2$.

\subsection{The detector acceptance}
The detector covers the full solid angle and, to a good
approximation, the acceptance is flat over the Dalitz plot. On the
other hand, some of the cuts introduce small biases and the exact
shape of the acceptance needs to be known. The acceptance was
determined from the reconstruction efficiency of $D_s$ production
and decay and the phase space which is smeared out by the finite
detector resolution. The reconstruction efficiency was derived by
the E791 collaboration using a full Monte Carlo simulation, from the
$\pi-N$ interaction to the digitalization and selection of the
events. The generated events had the nominal $D_s$ mass, the mass
distribution of reconstructed events was compatible with the
distribution shown in Fig. \ref{massdistr}a. For the analysis
presented here, this distribution was divided into 100 slices; then
events were generated, for each slice, simulating $D^+_s\to\pi^-
\pi^+ \pi^+$ decays uniformly spread over the phase space. The
summation of all Dalitz plot multiplied with the reconstruction
efficiency gave the acceptance $Acc(s_{12},s_{13})$ as presented in
Fig. \ref{massdistr}b ($s_{12}=m^2_{\pi^{^-}\pi^+_{(1)}}$,
$s_{13}=m^2_{\pi^{^-}\pi^+_{(2)}}$). In this way, edge bins are
properly taking into account.

\subsection{The background}
The fit  in Fig. \ref{massdistr}a assigned 568 events to $D_s$
decays and 280 events to the background. The latter was studied
using Monte Carlo simulations. The most prominent background
contributions stem from \begin{enumerate} \item a general
combinatorial background, \item $D^+ \to K^-\pi^+\pi^+$ decays where
a $K^+$ is wrongly interpreted as a $\pi^+$, \item $D_s^+ \to
\eta^{\prime} \pi^+$ decays followed by $\eta'\to
\rho^0(770)\gamma$, $\rho^0(770)$ $\to\pi^+\pi^-$ with a missing
$\gamma$, \item $D^0\to K^-\pi^+$ decays with an extra track (mostly
from the primary vertex). \end{enumerate} Amount and shape of the
background is determined using both Monte Carlo simulations and
data. The contribution of the $D_s^+ \to \eta' \pi^+$ background is
negligible. The sum of contributions 1, 2, and 3 above, $Bg_1$,
populates uniformly the $D_s^+\to\pi^-\pi^+\pi^+$ Dalitz plot
region. In the observed Dalitz plot, its distribution is
proportional to the acceptance. This latter background represents
$88\pm 2$\% of the total background contribution to the Dalitz plot.
The shape of the background from $D^0\to K^-\pi^+$ decays, $Bg_2$
representing $12\pm2$\% of the total background, is taken from
simulations. It can be described analytically, the function is
reproduced in the second part of the r.h.s. of eq. (\ref{1.1}). The
total background, $Bg=Bg_1+Bg_2$, is a function of the (symmetrized)
Dalitz plot variables: \begin{eqnarray}
\label{1.1} Bg(s_{12},s_{13})\ =\ 1.103\cdot Acc(s_{12},s_{13}) +
\phantom{rrrrrrrr} \\ \phantom{rrrrrrrr} 0.0065\cdot
\left(g_{D_0}(s_{12},s_{13})+g_{D_0}(s_{13},s_{12})\right)/2 \ ,\nn
\end{eqnarray}
\begin{eqnarray}
{\rm where } \ \phantom{rrr}
g_{D_0}(s_{12},s_{13})\ = \phantom{rrrrrrrrrrrrrrrrrrrrr} \nn\\
\phantom{rrrrrrrr}\ 7.2343\, e^{-0.5((0.25-s_{12})/0.2)^2+((3.15-s_{13})/0.11)^2)}
\nonumber \end{eqnarray}

\begin{figure}[ph]
\begin{center}
\begin{tabular}{ccc}
\hspace{-3mm}\includegraphics[width=0.26\textwidth]{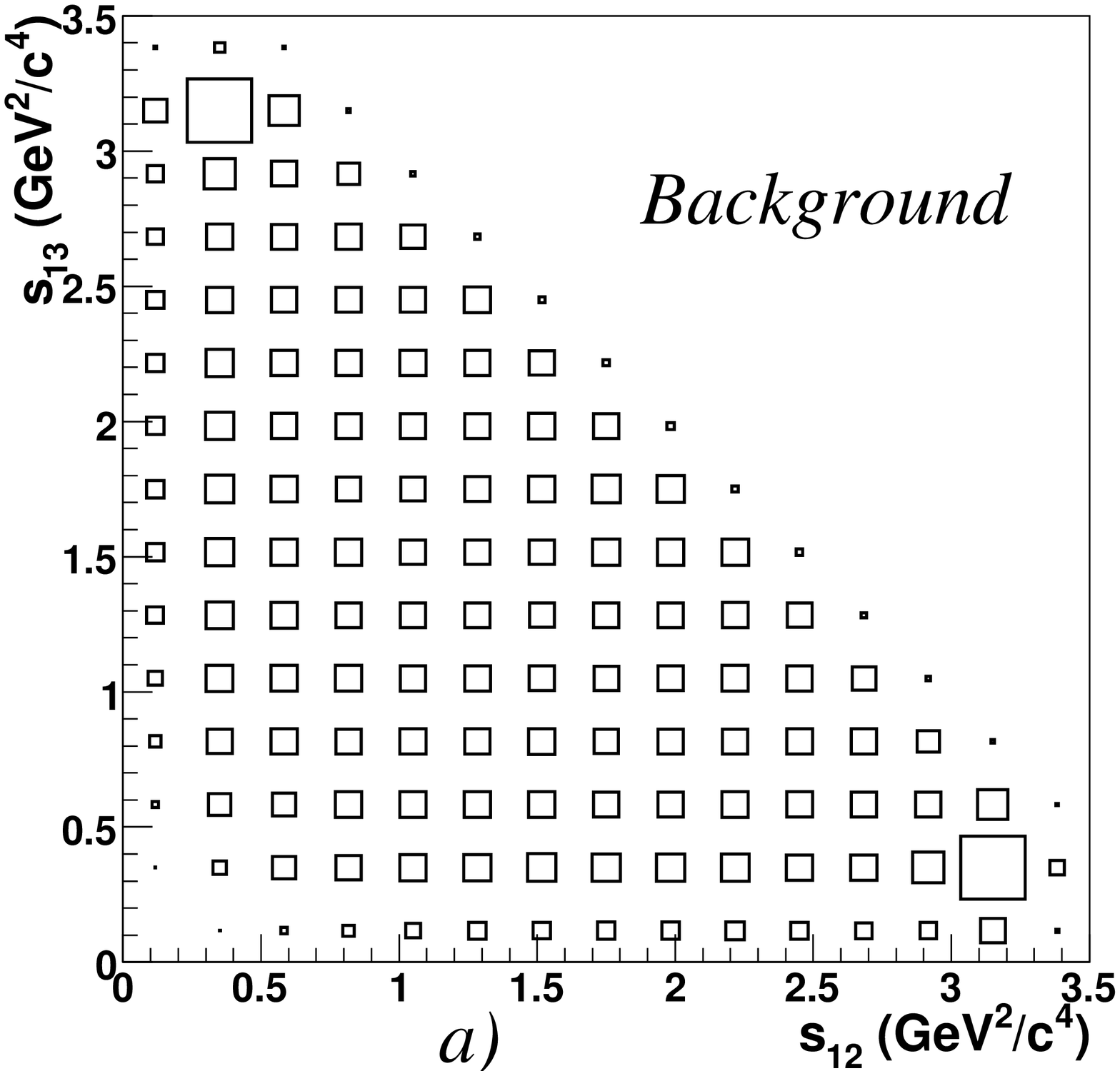}&
\hspace{-5mm}\includegraphics[width=0.26\textwidth]{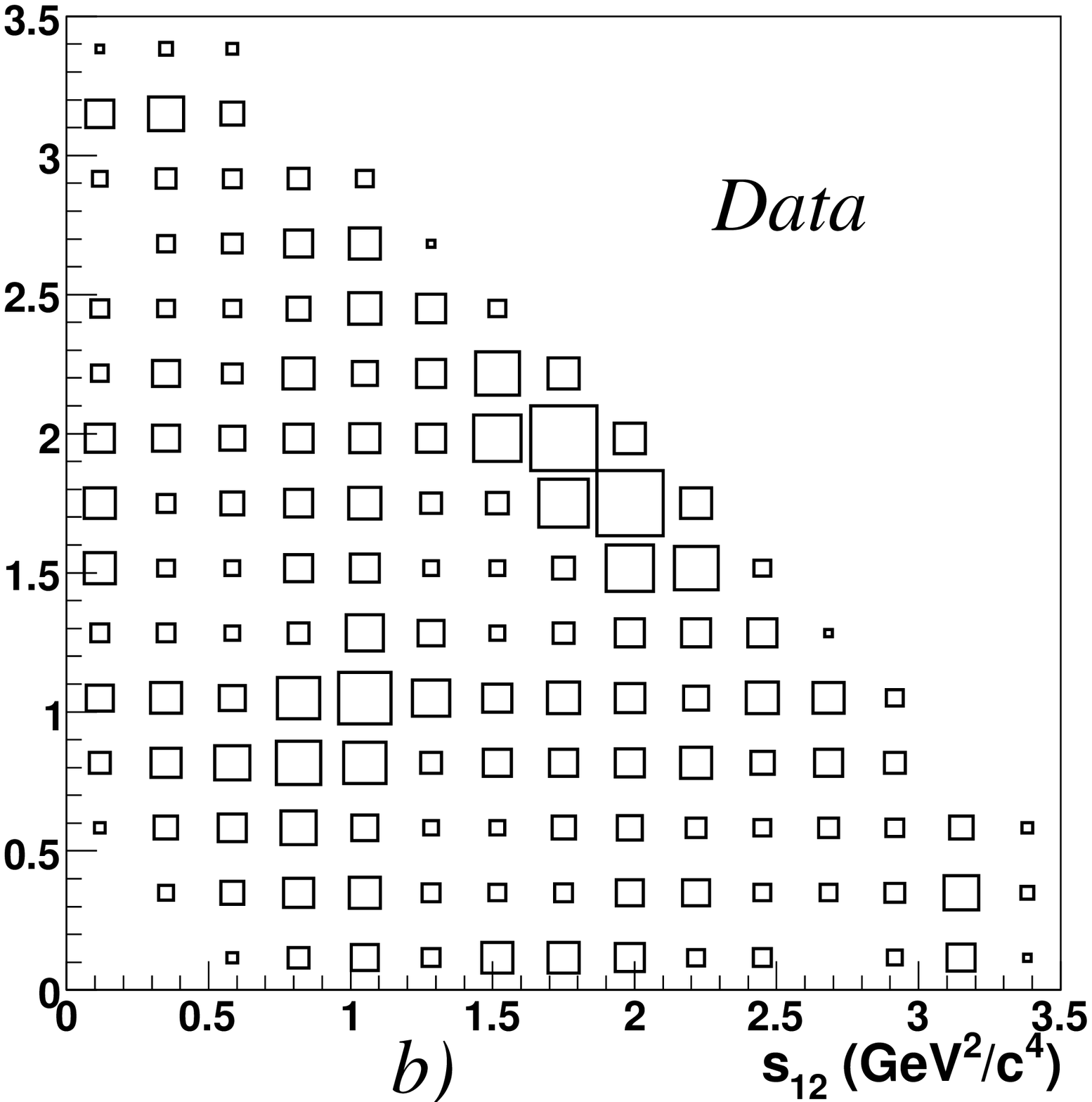}
\end{tabular}
\end{center}
\caption{\label{fig:data}a: The background Dalitz plot.
 b: The $D_s^+ \to \pi^- \pi^+ \pi^+$ Dalitz
plot. Since there are two identical particles, the plot is
symmetrized. } \vspace{-5mm}\end{figure}

After summation over the Dalitz plot, the background $Bg_1$ ($Bg_2$)
represents 246 (34) events. The Dalitz plot distribution of the
total background is shown in Fig. \ref{fig:data}a.

\subsection{The \boldmath$D_s^+ \to
\pi^- \pi^+ \pi^+$\unboldmath\ Dalitz plot}
 The symmetrized Dalitz plot for the $D^+_s$ candidates shown in Fig.
\ref{fig:data}b is reproduced from \cite{Aitala:2000xt}.  Since the two
$\pi^+$ are identical, the Dalitz plot is symmetrized. To avoid
problems with statistical errors, all entries receive a weight
$\frac 1 2$.

The most striking features of the Dalitz plot are the narrow
horizontal and vertical bands just below $s_{12} \simeq  s_{13}
\simeq $1 GeV$^2/c^4$ and a complex structure at 2\,GeV$^2/c^4$. The
bands correspond to $f_0(980) \pi^+_{(1)}$ decays, with $f_0(980)$
$\to \pi^+_{(2)}\pi^-$. The two $\pi^+$ are identical; two
interfering amplitudes contribute to the final state. The structure
at 2\,GeV$^2/c^4$ contains contributions from $f_2(1270)\pi^+$, from
$\rho^0(1460)$ $\pi^+$, and from $\pi^+$ recoiling against scalar
intensity. The clarification of this structure is the prime aim of
this paper.

\section{\label{PWA}Partial wave analysis}
\subsection{The fit function}
The partial wave analysis describes the Dalitz plot of Fig.
\ref{fig:data}b by a summation over possible reaction mechanisms for
$D_s^+\to 2\pi^+\pi^-$ decays plus background contributions. The
different reaction mechanisms can interfere, hence they are
represented by amplitudes. We consider the following reactions:
\begin{enumerate}
\item A $\pi^+$ recoiling against $\pi^+\pi^-$ in $S$-wave
\item A $\pi^+$ recoiling against $\pi^+\pi^-$ in $P$-wave
\item A $\pi^+$ recoiling against $\pi^+\pi^-$ in $D$-wave
\end{enumerate}
The $D$-wave is always described by a $f_2(1270)$ Breit-Wigner
amplitude. The $S$- and $P$-waves are alternatively represented by a
sum of two-channel Breit-Wigner amplitudes or by $K$-matrices.
Isotensor interactions are not included.

Due to Bose symmetry, the amplitudes are symmetrized with respect to
the exchange of the two $\pi^+$ mesons. We use the following notations:
$P=k_1+k_2+k_3$ is the momentum of the $D^+_s$ meson ($P^2=s$); $k_1$,
$k_2$ and $k_3$ are the pion momenta of the $\pi^-$ and of the two
$\pi^+$. Obviously, $k^2_1=k^2_2=k^2_3=m^2_\pi$ holds.
The amplitude depends of two invariant energy variables: \be
s_{12}=(k_1+k_2)^2 \ {\rm  and} \ s_{13}=(k_1+k_3)^2. \ee
The total amplitude includes contributions of different partial
waves ($S$-, $P$- and $D$- waves); it can be written in the form
 \be {\cal A}_{tot}(s_{12},s_{13})\ =&\\
&\hspace{-15mm} \sum_\alpha
\left[Z(k_1,k_3){\cal A}_\alpha(s_{12}) + Z(k_1,k_2){\cal
A}_\alpha(s_{13})\right] \nonumber \ee where $Z$ is a function referring
to the spin-orbital angular momentum structure of the two mesons
emerging from the decays of an intermediate state. For scalar, vector
and tensor waves, they have the following form ($i=2,3$):
\be
Z_S(k_1,k_i)=1,\qquad&\qquad Z_V(k_1,k_i)=z_{1i}, \\
Z_T(k_1,k_i)=\frac{1}{2}(3z^2_{1i}-1), \nn \quad&
\quad z_{1i}\ =\ \frac{k_{10}k_{30}-(k_1k_i)}{|\vec k_1||\vec
k_i|}.\nn
\ee

The amplitude contains the summation over the different isobars.
Each isobar depends on free fit parameters like masses, widths and
coupling constants.

The intensity observed in the Dalitz plot is proportional to the
squared reaction amplitude, to the phase space density $\Phi$ for $D_s$
decays and the background, and to the detector acceptance $Acc$.

\begin{eqnarray}
\label{1.4}
N(s_{12},s_{13})\ =\ |{\cal A}_{tot}(s_{12},s_{13})|^2 \cdot
Acc(s_{12},s_{13}) \Phi_{D_s} \phantom{rrr}\\
 + \alpha\cdot (Bg_1(1-\beta)+\beta Bg_2) \Phi_{Bg}\nn
\end{eqnarray}
The two background functions are given in eq. (\ref{1.1}). Note that
the background definition includes the acceptance.

For the amplitude ${\cal A}_{tot}(s_{12},s_{13})$ we use different
approa\-ches. The scalar amplitude contains a choice of a series of
poles for $\sigma(500)$, $f_0(980)$, $f_0(1370)$, $f_0(1500)$, and
$f_0(1710)$. The vector amplitude comprises up to three poles at 770
MeV/c$^2$, 1460\,MeV/c$^2$ and 1740\,MeV/c$^2$. The tensor wave is
described by a Breit-Wigner amplitude for the $f_2(1270)$ meson. In
most cases, we use fixed masses and widths.

As dynamical function describing meson resonances, we use $K$-matrices
or relativistic Breit-Wigner amplitudes.
\begin{enumerate}
\item In a first approach, we use $K$-matrices to describe scalar
resonances. For vector mesons, either a $K$-matrix or Breit-Wigner
functions are introduced.
\item In the second method, the $f_0(980)$ is represented by the
Flatt\'e formula. The total amplitude is formed as a sum of Flatt\'e
and Breit-Wigner amplitudes.
\end{enumerate}
\subsection{Breit-Wigner formalism.}
\subsubsection{Breit-Wigner amplitudes} The use of Breit-Wigner amplitudes offers
large flexibility and good control of the fit ingredients. The price
is the violation of unitarity when two Breit-Wigner amplitudes
overlap.

In the Breit-Wigner approach, the dynamical amplitude ${\cal
A}_{tot}(s_{12},s_{13})$ is given by a sum of relativistic
Breit-Wigner amplitudes which can be written in the form
 \be &{\cal A}_R= & BW_R(s)\ =\
\frac{\lambda_R^{\pi\pi}}{M_R^2-s-i\Gamma M_R} \nn \\ && \Gamma\ =\
\Gamma_R\left(\frac{p}{p_R}\right)^{2L}
\left(\frac{\rho}{\rho_R}\right)B'_L(p,p_R)
\ee
where $M_R$ is the mass and $\Gamma_R$ the partial width of the
resonance $R$. The complex number
$\lambda_R^{\pi\pi}=\Lambda_{D^+_s\pi}g_{\pi\pi}$ is given by the
product of the production amplitude $\Lambda_{D^+_s\pi}$ of the
resonance $R$ in $D_s$ decays, and the coupling constant
$g_{\pi\pi}$ for its decay into $\pi\pi$. The (running) decay
momentum is $p=\frac{1}{2}\sqrt{(s-4m^2_\pi)}$ while the decay
momentum at the nominal mass of the resonance is given by
$p_R=\frac{1}{2}\sqrt{(M^2_R-4m^2_\pi)}$. The phase space is written
as $\rho(s) = \sqrt{(s-4m_\pi^2)/s}$ and
$\rho_R(M^2_R)=\sqrt{(M^2_R-4m_\pi^2)/M^2_R}$, respectively.

For the three lowest orbital angular momenta, the Blatt-Weisskopf
factors $B'_L$ have the form
\be
\label{blt-weis}
&\hspace{-10mm}B'_0(p,p_R)\ =\ 1
&\hspace{-4mm}B'_1(p,p_R)\ =\ \sqrt{\frac{1+z_R}{1+z}} \nn \\
&\hspace{10mm}B'_2(p,p_R)\ =\
\sqrt{\frac{(z_R-3)^2+9z_R}{(z-3)^2+9z}}
\ee
where $z=(|p|d)^2$ is the meson radius. In the fits, the scale
parameter $d$ is restricted to $d\le 0.8\,fm$.

\subsubsection{The Flatt\'e parametrisation}
The $f_0(980)$ is taken into account using the Flatt\'e
parame\-trisation \be BW_{f_0(980)}(s)\ =\
\frac{\lambda_{f_0(980)}^{(\pi\pi)} +
i\rho_{K}\lambda_{f_0(980)}^{(K\bar K)}} {M_{f_0(980)}^2-s -
ig^2_\pi\rho_{\pi\pi} - ig^2_K\rho_{K\bar K}} \ee where
$M_{f_0(980)}$ and $\Gamma_{f_0(980)}$ are the $f_0(980)$ mass and
width, $\lambda_{f_0(980)}^{(\pi\pi)}$ and
$\lambda_{f_0(980)}^{(K\bar K)}$ are complex numbers. The first
number ($\lambda^{\pi\pi}$) is the same as the one used in the
Breit-Wigner parametrization. The second expression ($\lambda^{K\bar
K}$) refers to non-resonant two-kaon production and rescattering
into $\pi\pi$. The coupling constants for $f_0(980)$ decay into
$\pi\pi$ and $K\bar K$ are denoted as $g_\pi$ and $g_K$; the
two-particle phase space is written as
$\rho_{\pi\pi}=\sqrt{(s-4m^2_\pi)/s}$ and $\rho_{K\bar
K}=\sqrt{(s-4m^2_K)/s}$.

\subsubsection{Parameters of resonances}

The $T$-matrix poles of the most important resonances are listed in
Tables \ref{mg-vt} and \ref{mg}. In most of the fits described
below, masses, widths and some coupling constants are frozen.
\begin{table}[ph]
\vspace{-2mm} \caption{\label{mg-vt}The $\rho_0(1450)$ and
$f_2(1270)$ mass and width ($M+i\frac{\Gamma}{2}$) }
\renewcommand{\arraystretch}{1.3}
\bc \begin{tabular}{cc} \hline\hline
$\rho(1450)$ &$f_2(1270)$  \\
$1460+i\,150$ & $1275+i\,92.5$ \\
\hline\hline \end{tabular} \ec
\renewcommand{\arraystretch}{1.0}
\vspace{-8mm}
\end{table}
\begin{table}[ph] \caption{\label{mg}Masses and widths
($M+i\frac{\Gamma}{2}$ in GeV/c$^2$) of scalar resonances used in
fits.}
\renewcommand{\arraystretch}{1.3} \bc \begin{tabular}{cccccc}
\hline\hline $\sigma(500)$ &$f_0(1370)$&$f_0(1500)$ &$f_0(1710)$ \\
$0.55+i\,200$&$1.32+i\,155$&$1.490+i\,58.0$&$1.769+i\,170$\\
\hline\hline \vspace{-2mm}\end{tabular} \ec
\renewcommand{\arraystretch}{1.0} \end{table}

The $f_0(980)$ is described by the Flatt\'e formula; the following
parameters are used \cite{Anisovich:2002ij}:
 \be
&&M_{f_0(980)}=1.023\, {\rm GeV};
\\ \nonumber
&&g^2_{\pi}=(0.15-0.20)\,{\rm GeV}^2; \quad
\frac{g^2_{K}}{g^2_{\pi}}=2.4
 \ee
In all cases, we keep the names of Particle Data Group even though
we assign different masses to the particles.
\subsection{\label{K}K-matrix approach}
\subsubsection{The formalism}

As reference fit, we use a $K$-matrix in the $P$-vector approach of
Anisovich and Sarantsev \cite{Anisovich:2002ij}, applied to the
analysis of $D$ and $D_s$ decays in \cite{Link:2003gb}. The
$S$-wave amplitude has the form:
\be
\label{kmeq}
{\cal A}_i(s_{12}) \ =\
\frac{P_j(s_{12})}{I-i\rho_{ij} K_{ij}(s_{12})} \ee\be \nn
{\rm  where} \phantom{rrrrrrr}
&P_j(s_{12})\ =&\ \sum_\alpha \frac{\Lambda_\alpha
g^\alpha_j}{M^2_\alpha-s_{12}}+d_j \phantom{rrrrrr}
\ee\be\nn
 K_{ij}(s_{12}) =
  \left\{
    \sum_\alpha \frac{g^{\alpha}_i g^{\alpha}_j}{M^2_{\alpha}-s_{12}}
    + f^{\scatt}_{ij}\frac{1\,\mathrm{GeV}^2 - s_0^{\scatt}}{s_{12}-s_0^{\scatt}}
  \right\} \times \nn\\  \nn
  \frac{(s_{12}-s_A m^2_\pi/2)(1-s_{A0})}{(s_{12}-s_{A0})} + c_{ij}
\phantom{rrrrrrr} \ee
The production of a two-meson intermediate
state $j$ is enhanced due to formation of resonances. The summation in
(\ref{kmeq}) extends over all considered scalar resonances $\alpha$
with mass $M_\alpha$. The enhancement is proportional to their
production strengths $\Lambda_\alpha$ (for which we use $S_\alpha$,
$V_\alpha$, and $T_\alpha$ to denote the strength to produce scalar,
vector or tensor resonances), and their couplings $g^\alpha_j$ to the
decay channel $j$. The constants $d_j$ allow for nonresonant production
of that channel; however, this flexibility is not exploited in this
paper. The rescattering series is summed up in the $K$-matrix.
Transitions from unobserved channels $i$ into the observed channel $j$
by resonant rescattering are included, nonresonant meson-meson
scattering processes are taken into account by matrix elements
$c_{ij}$. The $\rho_{ij}$ form the phase space matrix elements: \be
\rho_{ij}=
\frac{1}{s}\sqrt{\left(s-(m_i+m_j)^2\right)\left(s-(m_i-m_j)^2\right)}
\ee
 The rescattering process $K\bar K\to \pi\pi$ is allowed below
the $K\bar K$ threshold; in this kinematical region, $\rho_{ij}$
becomes imaginary. The parameters $f^{\scatt}_{ij}$ and
$s_0^{\scatt}$ describe a slowly varying part of the $K$-matrix
elements. The masses $M_\alpha$ and the couplings $g^\alpha_j$,
$f^{\scatt}_{ij}$ are process-independent properties; they were
determined in \cite{Anisovich:2002ij} from a large number of data
sets and enter as fixed parameters into the analysis presented here.
The constants $\Lambda_\alpha$ and $d_j$ determine the dynamics of a
production process; they are free parameters to be determined from
the fits.

In a scattering situation, Chiral Perturbation Theory forces the
$\pi\pi$ amplitude to vanish at small energies. This is a
kinematical effect which is not enforced in production. The
suppression of the $\pi\pi$ scattering amplitude at small energies
is taken into account by the Adler-Weinberg zero,
$(s-s_Am^2_{\pi}/2)(1-s_{A0})/(s-s_{A0})$. We choose $s_A=1,
s_{A0}=-0.15$.

For higher partial waves, only the resonant terms are required to
describe data. The $K$-matrix now contains Blatt-Weisskopf barrier
factors $B_{L}$
\be\nn
 K_{ij}(s_{12}) =
    \sum_\alpha \frac{1}{B'_{L(ij)}}
\frac{g^{\alpha}_i
g^{\alpha}_j}{M^2_{\alpha}-s_{12}}\frac{1}{B'_{L(ij)}} \ee which
depend on the orbital angular momentum $L$ between the two mesons,
the c.m.s. momentum of the two mesons in the initial and final state
and one scale parameter. The Blatt-Weisskopf factors are given in
eq. (\ref{blt-weis}). The production vector for channel $P_j$ is
multiplied by one Blatt-Weisskopf factor taking into account the
angular momentum barrier for channel $j$.

In this paper, the $K$-matrix from  \cite{Anisovich:2002ij} is used for
the $S$-wave which included five resonances. These five resonances are
observed when the $T$-matrix poles of the scattering amplitude are
inspected. The $T$ matrix is defined by
\be T=(I-i
K\cdot\rho)^{-1}K\,. \ee
 The $T$-matrix poles depend on the $K$-matrix masses and the
couplings $g^\alpha_j$ and are thus independent of the production
process. The poles were identified with $f_0(980)$, $f_0(1370)$,
$f_0(1500)$, $f_0(1710)$ of the Particle Data Listings \cite{PDG}
plus a broad resonance $f_0(1470)$ which was interpreted as scalar
glueball.

$K$-matrix masses and coupling constants $g^{\alpha}_{j}$ and
$f^{\scatt}_{1j}$ are taken from \cite{Anisovich:2002ij};  the
values are reproduced in Table \ref{anisar_scal1} and
\ref{anisar_scal2} for two cases. In Table \ref{anisar_scal1}, a
solution is given which includes a pole for $f_0(1370)$; the
parameters in Table \ref{anisar_scal2} are optimized for the case
where $f_0(1370)$ is omitted. The $K$-matrix parameters reproduce
the $T$-matrix poles given in Tables \ref{mg-vt} and \ref{mg}.

\begin{table}[htb]
\caption{\label{anisar_scal1}$K$-matrix parameters for the scalar
isoscalar wave using 5 $K$-matrix poles. Masses and coupling
constants are in GeV. Only the $i=1$ $f^{\scatt}_{ij}$ terms are
listed since they are the only values relevant to the three-pion
decay. }
\bc\renewcommand{\arraystretch}{1.4}
$\begin{array}{cccccc}\hline\hline m_{\alpha} & g_{\pi \pi} & g_{K
\bar K} & g_{4 \pi} & g_{\eta \eta} & g_{\eta \eta' }\\ \hline
0.65100&0.22889&-0.55377& 0.00000&-0.39899&-0.34639 \\
1.20360&0.94128& 0.55095& 0.00000& 0.39065& 0.31503 \\
1.55817&0.36856& 0.23888& 0.55639& 0.18340& 0.18681 \\
1.21000&0.33650& 0.40907& 0.85679& 0.19906&-0.00984 \\
1.82206&0.18171&-0.17558&-0.79658&-0.00355& 0.22358 \\
\hline s_0^{\scatt} & f^{\scatt}_{11} & f^{\scatt}_{12} &
f^{\scatt}_{13} & f^{\scatt}_{14} & f^{\scatt}_{15} \\
-3.92637&0.23399&0.15044&-0.20545&0.32825&0.35412 \\
 \hline
\hline
\end{array}
$\renewcommand{\arraystretch}{1.0}\ec
\caption{\label{anisar_scal2}$K$-matrix parameters for the scalar
isoscalar wave using 4 $K$-matrix poles. The pole representing
$f_0(1370)$ was omitted resulting, in \cite{Anisovich:2002ij}, in a
increased $\chi^2$. Masses and coupling constants are in GeV. }
\bc\renewcommand{\arraystretch}{1.4} $\begin{array}{cccccc}\hline\hline
m_{\alpha} & g_{\pi \pi} & g_{K \bar K} & g_{4 \pi} & g_{\eta \eta} &
g_{\eta \eta' }\\ \hline 0.69566&0.66859&-0.71636&
0.00000&-0.10133&-0.17721 \\ 1.24458&0.87180& 0.59038& 0.00000&
0.38115& 0.39085 \\ 1.54280&0.36228& 0.24167& 0.50442& 0.15035& 0.45126
\\ 1.84274&0.07290&-0.10118&-0.30778& 0.03018&-0.17852      \\
\hline s_0^{\scatt} & f^{\scatt}_{11} & f^{\scatt}_{12} &
f^{\scatt}_{13} & f^{\scatt}_{14}& f^{\scatt}_{15} \\
-3.84414&0.54458&-0.10980&-0.55817&0.46870&-0.02877 \\
\hline
\hline
\end{array}
$\renewcommand{\arraystretch}{1.0}\ec
\end{table}

For the vector mesons, a $K$-matrix is used which contains $\rho(770)$,
$\rho(1450)$, and $\rho(1770)$. The $K$-matrix masses and coupling
constants are reproduced in Table \ref{anisar_vec}.
\begin{table}[pt]
\caption{\label{anisar_vec}$K$-matrix parameters for the vector
isovector wave. Masses and coupling constants are in GeV, $d$ is
given in $fm$. Only the $i=1$ $c_{ij}$ terms are listed since they
are the only values relevant to the three-pion decay.}
\bc\renewcommand{\arraystretch}{1.4}$\begin{array}{ccccc}\hline\hline
m_{\alpha}&d&g_{\pi\pi}&g_{4\pi}&g_{\pi\omega}\\
\hline
0.77873&0.34469& 0.69461&-0.50000& 0.00000 \\
1.50000&0.80000&-0.20000& 0.54000&-0.65181 \\
1.90000&0.80000& 0.18042&-0.71729&-0.30664 \\
\hline
\hline
\end{array}$\renewcommand{\arraystretch}{1.0}\ec
\end{table}

\subsection{Fit results}
\subsubsection{Study of vector and tensor contributions}
\begin{figure}[pb]
\vspace{-5mm}
\begin{center}
\begin{tabular}{cc}
             \includegraphics[width=0.265\textwidth]{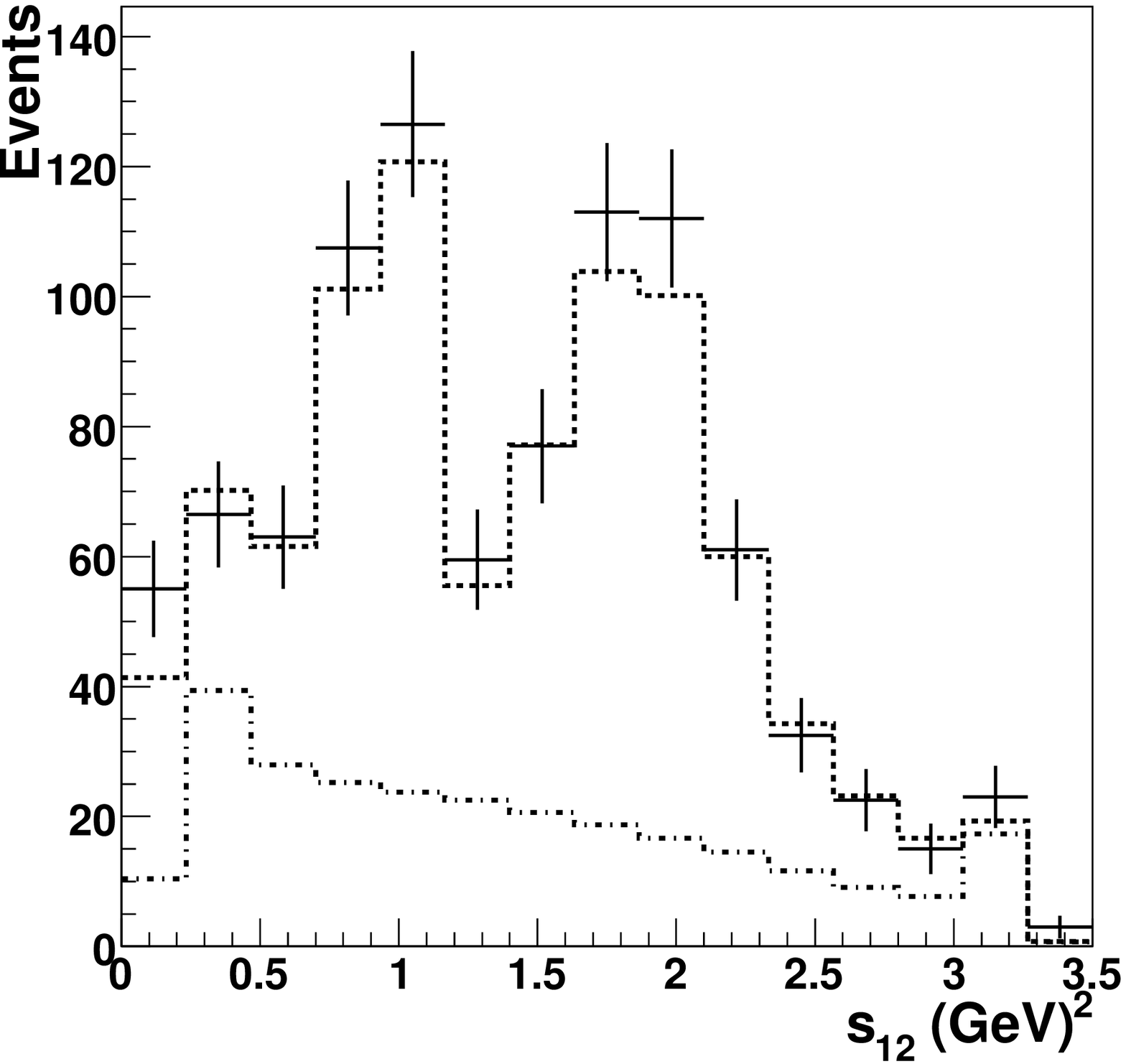}&
\hspace{-8mm}\includegraphics[width=0.265\textwidth]{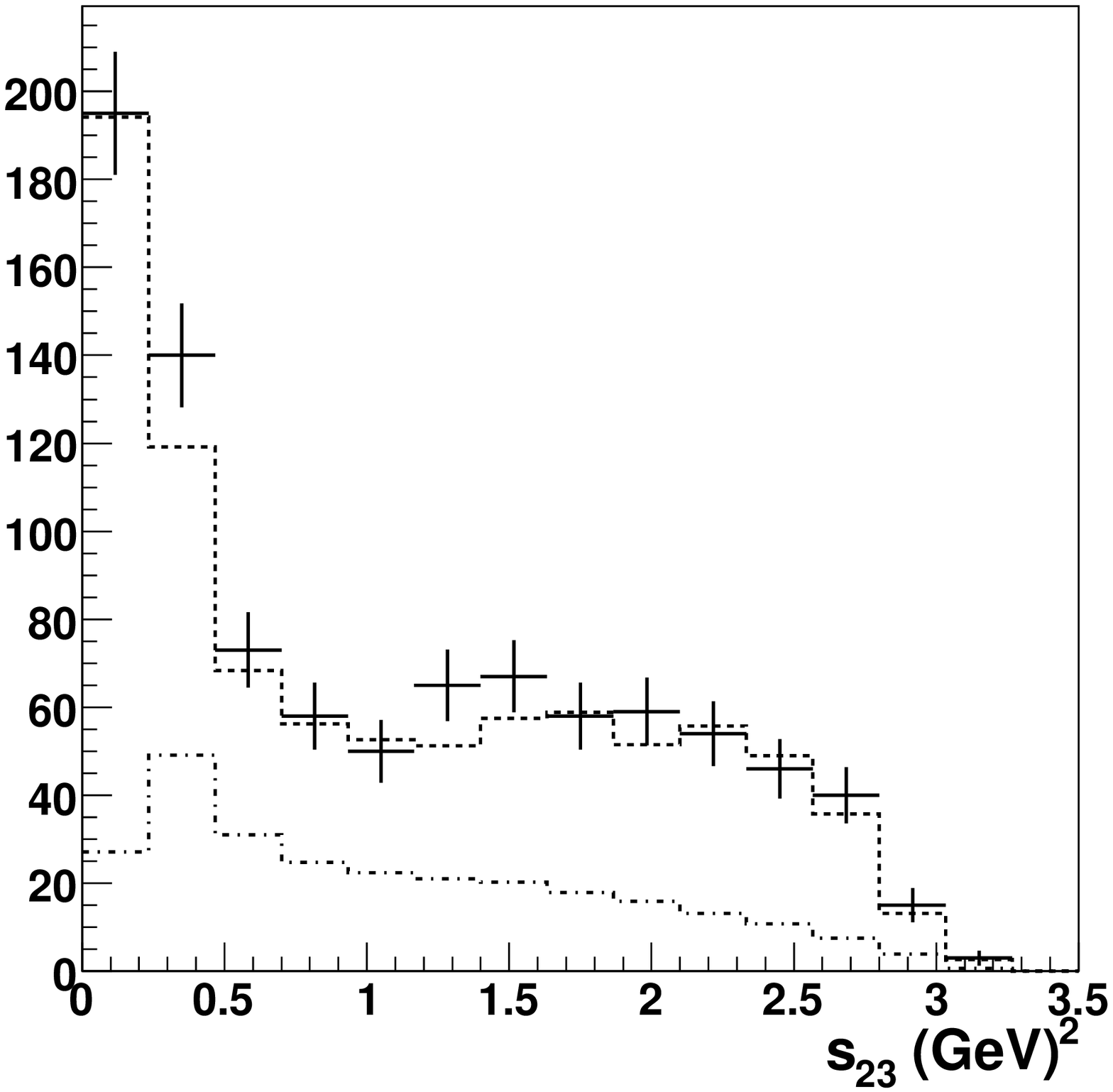}
\vspace{-5mm}
\end{tabular}
\end{center}
\caption{\label{sol1-dist}
The experimental mass distributions are
compared with the Reference fit (solution 1 of Table \ref{kresult-pf})
to $D_s^+\to\pi^+\pi^+\pi^-$ decays.}
\end{figure}

\begin{figure}[pt]
\begin{center}
\includegraphics[width=0.48\textwidth]{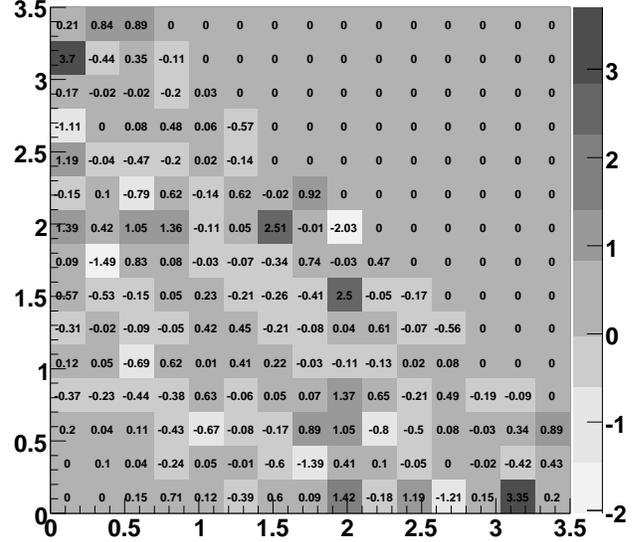}
\end{center}
\caption{\label{sol1-chi2}Reference fit (solution 1 of Table
\ref{kresult-pf}) to the data $D_s^+\to\pi^+\pi^+\pi^-$. The
$\chi^2$ contributions of the individual Dalitz plot cells are given
in a grey scale and in numbers. For data exceeding the fit, the
cells are dark and the $\chi^2$ is plotted with a positive sign.
When the fit exceeds data, the cells are white and $\chi^2$ is
plotted with a negative sign.} \end{figure}

Our fit strategy is as follows: we first describe the $S$-wave and
$P$-waves using the $K$-matrices with fixed pole structure as given
in Tables \ref{anisar_scal1} and \ref{anisar_vec}. The $f_2(1270)$
is included as Breit-Wigner resonance. This fit has 13 parameters to
describe the 77 independent cells. Hence there are 64 degrees
freedom. The $\chi^2$ of the fit is 63.49; the fit quality is very
satisfactory. Figs. \ref{sol1-dist} and \ref{sol1-chi2} show mass
square distributions and, respectively, the $\chi^2$ per Dalitz plot
cell. A negative sign is attached to the $\chi^2$ whenever the fit
result exceeds data.

Results of this fit and the $\chi^2$ achieved is given
in Table \ref{kresult-pf} as solution 1 (Sol. 1). The (complex)
production amplitudes are given without errors; the statistical errors
are small in comparison to the spread of results when the fit
hypothesis is varied.

In a next step we explore the role of vector and tensor mesons
maintaining the full flexibility of the $S$-wave description. First,
we replace the $K$-matrix by three Breit-Wigner amplitudes. A
slightly improved $\chi^2$ is obtained, $\delta\chi^2=-0.85$ (Sol.
2). The results do not change significantly when $\rho(770)$ is
removed from the fit. Without $\rho(770)$, $\chi^2$ increases by
0.21 compared to solution 2, the number of parameters by 2 (Sol. 3).
Removing $\rho(1770)$ changes $\chi^2$ by 1.75 (Sol. 4), and
removing both, $\rho(770)$ and $\rho(1770)$, leads to $\chi^2=64.41$
(Sol. 5), an increase in $\chi^2$ by about 2 units while four
parameters are spared. We conclude that $\rho(1460)$ is sufficient
to describe the contribution of vector mesons to the data. However,
if the latter resonance or the $f_2(1270)$ is taken out of the fit,
$\chi^2$ increases by 20 or more. Both these resonances are required
to get a good fit. A further small improvement is achieved when mass
and width of the $\rho$ and $f_2$ resonances are allowed to vary
freely. The resulting values remain compatible with PDG values. Both
are included with fixed values for mass and width in all subsequent
fits (see Table \ref{mg-vt}). We note that some fits prefer to split
$\rho(1450)$ into a $\rho(1250)$ and a $\rho(1500)$ even though the
statistical evidence for the split $\rho(1450)$ is weak
($\delta\chi^2=4$ for two more parameters).

 \begin{table}[pt] \vspace{-3mm} \caption{\label{kresult-pf}Fit
parameters for five different solutions exploring the role of vector
and tensor resonances. The full $\pi\pi$ $S$-wave is always
included. $S_1$ and $S_2$ interfere to make the $f_0(980)$,
$S_3=f_0(1370)$, $S_4=f_0(1500)$, $S_5=f_0(1710)$. There are up to
three vector resonances, $V_1=\rho(770)$, $V_2=\rho(1450)$, and
$V_3=\rho(1770)$, and the  tensor $T_1=f_2(1270)$. Given are the
complex production strengths $\Lambda_{\alpha}$ of eq. (\ref{kmeq})
and the $\chi^2$ of the fit.}
{\scriptsize\bc\renewcommand{\arraystretch}{1.3}\begin{tabular}{cccccc}
\hline\hline
&Sol.1\phantom{rrr}&Sol.2\phantom{rrr}&Sol.3\phantom{rrr}&Sol.4\phantom{rrr}&Sol.5\phantom{rrr}\\
\hline
\hspace{-2.5mm}$S_1$\hspace{-4mm}&\hspace{-3.5mm}8.10-i5.03\hspace{-3.5mm}&\hspace{-3.5mm}8.61-i3.80\hspace{-3.5mm}&\hspace{-3.5mm}8.70-i3.68\hspace{-3.5mm}&\hspace{-3.5mm}6.68-i6.37\hspace{-3.5mm}&\hspace{-3.5mm}6.78-i6.31\hspace{-3.5mm}\\
\hspace{-2.5mm}$S_2$\hspace{-4mm}&\hspace{-3.5mm}-7.83+i2.25\hspace{-3.5mm}&\hspace{-3.5mm}-7.71+i1.76\hspace{-3.5mm}&\hspace{-3.5mm}-7.73+i1.64\hspace{-3.5mm}&\hspace{-3.5mm}-7.54+i3.36\hspace{-3.5mm}&\hspace{-3.5mm}-7.55 +i3.29\hspace{-3.5mm} \\
\hspace{-2.5mm}$S_3$\hspace{-4mm}&\hspace{-3.5mm}-1.81+i5.46\hspace{-3.5mm}&\hspace{-3.5mm}-2.34+i6.85\hspace{-3.5mm}&\hspace{-3.5mm}-2.54+i6.49\hspace{-3.5mm}&\hspace{-3.5mm}-2.01+i9.66\hspace{-3.5mm}&\hspace{-3.5mm}-1.82+i9.28\hspace{-3.5mm}\\
\hspace{-2.5mm}$S_4$\hspace{-4mm}&\hspace{-3.5mm}-2.22+i5.95\hspace{-3.5mm}&\hspace{-3.5mm}-1.78+i7.53\hspace{-3.5mm}&\hspace{-3.5mm}-2.55+i6.94\hspace{-3.5mm}&\hspace{-3.5mm}-1.10+i11.69\hspace{-3.5mm}&\hspace{-3.5mm}-0.75+i11.21\hspace{-3.5mm}\\
\hspace{-2.5mm}$S_5$\hspace{-4mm}&\hspace{-3.5mm}4.94-i6.13\hspace{-3.5mm}&\hspace{-3.5mm} 6.82-i7.85\hspace{-3.5mm}&\hspace{-3.5mm}6.91-i7.10\hspace{-3.5mm}&\hspace{-3.5mm}12.70-i10.78\hspace{-3.5mm}&\hspace{-3.5mm}11.63-i10.70\hspace{-3.5mm}\\
\hspace{-2.5mm}$V_1$\hspace{-4mm}&\hspace{-3.5mm}0.10+i0.14\hspace{-3.5mm}&\hspace{-3.5mm}-0.06-i0.12\hspace{-3.5mm}&---\phantom{rrr}                        &\hspace{-3.5mm}0.04-i0.04\hspace{-3.5mm}&---\phantom{rrr}\\
\hspace{-2.5mm}$V_2$\hspace{-4mm}&\hspace{-3.5mm}3.00-i15.03\hspace{-3.5mm}&\hspace{-3.5mm}2.86-i0.23\hspace{-3.5mm}&\hspace{-3.5mm} 2.84-i0.29\hspace{-3.5mm}&\hspace{-3.5mm} 2.22 +i0.75\hspace{-3.5mm}&\hspace{-3.5mm}2.15+i0.85\hspace{-3.5mm}  \\
\hspace{-2.5mm}$V_3$\hspace{-4mm}&\hspace{-3.5mm}7.98-i7.76\hspace{-3.5mm}&\hspace{-3.5mm}-3.08+i0.61\hspace{-3.5mm}&\hspace{-3.5mm}-2.90+i0.69\hspace{-3.5mm}&---\phantom{rrr}&---\phantom{rrr}\\
\hspace{-2.5mm}$T_1$\hspace{-4mm}&\hspace{-3.5mm}-5.43+i0.00\hspace{-3.5mm}&\hspace{-3.5mm}-5.30+i0.00\hspace{-3.5mm}&\hspace{-3.5mm}-5.31+i0.00\hspace{-3.5mm}&\hspace{-3.5mm}-5.623 +i0.000 \hspace{-3.5mm}&\hspace{-3.5mm}-5.58+i0.00\hspace{-3.5mm}\\
\hline
$\chi^2$                         &63.49\phantom{rrr}                          &62.54\phantom{rrr}                          &62.75\phantom{rrr}                          &64.29\phantom{rrr}                           &64.41\phantom{rrr}                            \\
\hline\hline \end{tabular}
\vspace{-2mm}
\renewcommand{\arraystretch}{1.0}\ec}
\end{table}

\subsubsection{Study of the scalar partial
wave} Table \ref{kresult-s} presents a series of fits in which
amplitudes contributing to the $\pi\pi$ $S$-wave are explored. For
convenience of the reader we reproduce solution 5 of Table
\ref{kresult-pf} as solution 6 in Table \ref{kresult-s}. We then add
a constant $N_1=d_{\pi\pi}$ describing direct $\pi\pi$ production.
This gives an improvement of 2.67 in $\chi^2$ (Sol. 7). Direct
production of a $K\bar K$ pair ($N_2$) and subsequent rescattering
into $\pi\pi$ is unimportant:  $\chi^2$ changes by 0.32 units only
(Sol. 8).

\begin{table}[pt]
\caption{\label{kresult-s}Study of the scalar wave in the $K$-matrix
approach. See caption of Table \ref{kresult-pf}.}
{\scriptsize\bc\renewcommand{\arraystretch}{1.3}\begin{tabular}{cccccc}
\hline\hline
&Sol.6\phantom{rrr}&Sol.7\phantom{rrr}&Sol.8\phantom{rrr}&Sol.9\phantom{rrr}&Sol.10\phantom{rrr}\\
\hline
\hspace{-2.5mm}$S_1$\hspace{-4mm}&\hspace{-3.5mm}6.78-i6.31\hspace{-3.5mm}&\hspace{-3.5mm}6.50-i4.39\hspace{-3.5mm}&\hspace{-3.5mm}6.81-i4.22\hspace{-3.5mm}&\hspace{-3.5mm}9.05-i1.07\hspace{-3.5mm}&\hspace{-3.5mm}8.42-i2.19\hspace{-3.5mm}\\
\hspace{-2.5mm}$S_2$\hspace{-4mm}&\hspace{-3.5mm}-7.55+i3.23\hspace{-3.5mm}&\hspace{-3.5mm}-7.78+i2.31\hspace{-3.5mm}&\hspace{-3.5mm}-5.39-i0.13\hspace{-3.5mm}&\hspace{-3.5mm}-8.73+i0.53\hspace{-3.5mm}&\hspace{-3.5mm}-9.71+i0.36\hspace{-3.5mm}\\
\hspace{-2.5mm}$S_3$\hspace{-4mm}&\hspace{-3.5mm}-1.82+i9.28\hspace{-3.5mm}&\hspace{-3.5mm}0.88+i4.43\hspace{-3.5mm}&\hspace{-3.5mm}0.45-i1.418\hspace{-3.5mm}&\hspace{-3.5mm}-1.79+i1.49\hspace{-3.5mm}&\hspace{-3.5mm}-2.51-i0.62\hspace{-3.5mm} \\
\hspace{-2.5mm}$S_4$\hspace{-4mm}&\hspace{-3.5mm}-0.75+i11.21\hspace{-3.5mm}&\hspace{-3.5mm}2.68+i4.32\hspace{-3.5mm}&\hspace{-3.5mm}-1.18-i2.64\hspace{-3.5mm}&---\phantom{rrr}&---\phantom{rrr}\\
\hspace{-2.5mm}$S_5$\hspace{-4mm}&\hspace{-3.5mm}11.63-i10.70\hspace{-3.5mm}&\hspace{-3.5mm} 2.68-i11.03\hspace{-3.5mm}&\hspace{-3.5mm}-0.02 -i3.52\hspace{-3.5mm}&\hspace{-3.5mm}10.97-i3.34\hspace{-3.5mm}&\hspace{-3.5mm}4.73-i7.80\hspace{-3.5mm}\\
\hspace{-2.5mm}$N_1$\hspace{-4mm}&---\phantom{rrr}&\hspace{-3.5mm}-1.71+i4.12\hspace{-3.5mm}&\hspace{-3.5mm}-2.41+i6.10\hspace{-3.5mm}&---\phantom{rrr}&\hspace{-3.5mm}0.05+i4.69\hspace{-3.5mm}\\
\hspace{-2.5mm}$N_2$\hspace{-4mm}&---\phantom{rrr}&---\phantom{rrr}&\hspace{-3.5mm}-1.30+i12.76\hspace{-3.5mm}&---\phantom{rrr}&---\phantom{rrr}\\
\hspace{-2.5mm}$V_2$\hspace{-4mm}&\hspace{-3.5mm}2.15+i0.85\hspace{-3.5mm}&\hspace{-3.5mm}2.86+i0.29\hspace{-3.5mm}&\hspace{-3.5mm}2.80+i0.37\hspace{-3.5mm}&\hspace{-3.5mm}1.70+i0.20\hspace{-3.5mm}&\hspace{-3.5mm}2.58-i0.25\hspace{-3.5mm}\\
\hspace{-2.5mm}$T_1$\hspace{-4mm}&\hspace{-3.5mm}-5.58+i0.00\hspace{-3.5mm}&\hspace{-3.5mm}-5.07+i0.00\hspace{-3.5mm}&\hspace{-3.5mm}-4.92+i0.00\hspace{-3.5mm}&\hspace{-3.5mm}-5.39+i0.00\hspace{-3.5mm}&\hspace{-3.5mm}-5.05+i0.00\hspace{-3.5mm}\\
\hline
$\chi^2$                         &64.41\phantom{rrr}                           &61.74\phantom{rrr}                            &61.42\phantom{rrr}                            &70.85\phantom{rrr}                           &66.12\phantom{rrr}                           \\
\hline\hline \end{tabular}
\renewcommand{\arraystretch}{1.0}
\vspace{-5mm}
\ec}
\end{table}
We now take out $f_0(1370)$. This is an alternative solution
presented in \cite{Anisovich:2002ij}. In a fit to a large number of
reactions, the omission of $f_0(1370)$ resulted in a worse $\chi^2$.
The $K$-matrix parameters obtained in that fit are listed in Table
\ref{anisar_scal2}. In the fit to the $D_s^+\to\pi^+\pi^+\pi^-$
data, $\chi^2$ changes by 6.44 units when solutions 6 (with
parameter $N_1=0$) is compared to solution 9, and by 4.38 when
solution 7 with $N_1\ne 0$ is compared to solution 10, see Table
\ref{kresult-s}. Including $N_2$ in the fit with no $f_0(1370)$
gives again a marginal improvement.

In a next set of fits, we used the Flatt\'e parametrization to
describe the $f_0(980)$, and Breit-Wigner amplitudes for all other
resonances. The three states $f_0(980)$, $\rho(1450)$, and
$f_2(1270)$ are always included. For the $S$-wave, the maximal set
of resonances includes, apart from $f_0(980)$, the $\sigma(500)$,
$f_0(1370)$, $f_0(1500)$, and $f_0(1710)$. All masses and widths are
fixed to the $T$-matrix pole positions obtained in
\cite{Anisovich:2002ij}. Fit results are collected in Table
\ref{bwresult-1} and \ref{bwresult-2}.

A fit which includes all resonances (Sol. 11) yields the best
$\chi^2$, 62.47 for 16 parameters. The $f_0(980)$ is not sufficient
to describe the scalar intensity: when $\sigma(500)$, $f_0(1370)$,
$f_0(1500)$, and $f_0(1710)$ are removed, $\chi^2$ increases
dramatically to 110.18 (Sol. 12). The fit is unacceptable. We now
test the $\chi^2$ changes when only one of the three high-mass
resonances is included in the fit. Inclusion of $f_0(1500)$ gives
the smallest yield and the largest $\chi^2$ gain (Sol. 14). The
other two resonances (Sol. 13 and 15) need a large yield (to
describe intensity at the wrong mass) and bring a smaller gain in
$\chi^2$. We conclude that $f_0(1500)$ helps best to describe the
data efficiently.

Fit 16 reproduces the results of solution 11. Removing
$S_1=\sigma(500)$ changes $\chi^2$ by 3.06 units, and saves two
parameters (Sol. 17). The $\chi^2$ change is certainly not
sufficient to claim that $\sigma(500)$ is present but some
contribution can also not be excluded. Its integrated fractional
contribution to the Dalitz plot is estimated to $\sim 3$\%. In
solution 18, the $S_3=f_0(1370)$ is removed additionally, with leads
to a $\chi^2$ increase of 0.44 units while two parameters are
spared; the $f_0(1370)$ does not help to improve the fit.  If
$f_0(1500)$ is removed (Sol. 19), $\chi^2$ increases by 7.67. In
solution 20, $f_0(1710)$ is removed, and $\chi^2$ changes by 3.77
units. Again, $f_0(1500)$ has the largest impact on the fit quality.
The best $\chi^2$ per degree of freedom is achieved when the scalar
wave is described by $f_0(980)$, $f_0(1500)$, and $f_0(1710)$.

\begin{table}[pt]
\caption{\label{bwresult-1}Fits using Breit-Wigner amplitudes and
Flatt\'e parametrization for $f_0(980)$. $S_1=\sigma(500)$; the two
lines for $f_0(980)$, $S_2$ and $S_2'$, give the couplings
$\lambda^{\pi\pi}$ and $\lambda^{K\bar K}$. $S_3=f_0(1370)$,
$S_4=f_0(1500)$, $S_5=f_0(1710)$, $V_2=\rho(1450)$, $T_1=f_2(1270)$.
}
{\scriptsize\bc\renewcommand{\arraystretch}{1.3}\begin{tabular}{cccccc}
\hline\hline
&Sol.11\phantom{rrr}&Sol.12\phantom{rrr}&Sol.13\phantom{rrr}&Sol.14\phantom{rrr}&Sol.15\phantom{rrr}\\
\hline
\hspace{-2.5mm}$S_1 $\hspace{-4mm}&\hspace{-3.5mm} 0.73-i0.14\hspace{-3.5mm}&---\phantom{rrr}                         &---\phantom{rrr}                         &---\phantom{rrr}                         &---\phantom{rrr}                          \\
\hspace{-2.5mm}$S_2 $\hspace{-4mm}&\hspace{-3.5mm} 3.57+i5.18\hspace{-3.5mm}&\hspace{-3.5mm} 0.09+i5.35\hspace{-3.5mm}&\hspace{-3.5mm} 1.99+i3.03\hspace{-3.5mm}&\hspace{-3.5mm} 1.66+i4.48\hspace{-3.5mm}&\hspace{-3.5mm} 1.45+i5.79\hspace{-3.5mm} \\
\hspace{-2.5mm}$S_2'$\hspace{-4mm}&\hspace{-3.5mm} 2.53+i4.94\hspace{-3.5mm}&\hspace{-3.5mm}-4.62+i3.27\hspace{-3.5mm}&\hspace{-3.5mm}-3.59+i0.45\hspace{-3.5mm}&\hspace{-3.5mm}-2.78+i3.10\hspace{-3.5mm}&\hspace{-3.5mm}-1.73+i5.93\hspace{-3.5mm} \\
\hspace{-2.5mm}$S_3 $\hspace{-4mm}&\hspace{-3.5mm} 0.96-i0.58\hspace{-3.5mm}&---\phantom{rrr}                         &\hspace{-3.5mm}-2.13-i1.57\hspace{-3.5mm}&---\phantom{rrr}                         &---\phantom{rrr}                          \\
\hspace{-2.5mm}$S_4 $\hspace{-4mm}&\hspace{-3.5mm} 0.71-i0.71\hspace{-3.5mm}&---\phantom{rrr}                         &---\phantom{rrr}                         &\hspace{-3.5mm} 0.38-i0.98\hspace{-3.5mm}&---\phantom{rrr}                          \\
\hspace{-2.5mm}$S_5 $\hspace{-4mm}&\hspace{-3.5mm} 4.31-i1.36\hspace{-3.5mm}&---\phantom{rrr}                         &---\phantom{rrr}                         &---\phantom{rrr}                         &\hspace{-3.5mm} 3.76+i2.09\hspace{-3.5mm} \\
\hspace{-2.5mm}$V_2 $\hspace{-4mm}&\hspace{-3.5mm} 2.79+i1.89\hspace{-3.5mm}&\hspace{-3.5mm} 0.11+i3.02\hspace{-3.5mm}&\hspace{-3.5mm} 1.28+i1.96\hspace{-3.5mm}&\hspace{-3.5mm} 1.07+i2.46\hspace{-3.5mm}&\hspace{-3.5mm} 1.38+i2.94\hspace{-3.5mm} \\
\hspace{-2.5mm}$T_1 $\hspace{-4mm}&\hspace{-3.5mm}-5.39+i0.00\hspace{-3.5mm}&\hspace{-3.5mm}-5.77+i0.00\hspace{-3.5mm}&\hspace{-3.5mm}-5.89+i0.00\hspace{-3.5mm}&\hspace{-3.5mm}-5.45+i0.00\hspace{-3.5mm}&\hspace{-3.5mm}-5.40+i0.00\hspace{-3.5mm} \\
\hline
$\chi^2$                          &62.47\phantom{rrr}                       &110.18\phantom{rrr}                      &78.48\phantom{rrr}                       &70.56\phantom{rrr}                       &81.40\phantom{rrr}                        \\
\hline\hline
\end{tabular}
\renewcommand{\arraystretch}{1.0}
\vspace{-5mm}
\ec}
\end{table}
\begin{table}[pt]
\caption{\label{bwresult-2}Fits using Breit-Wigner amplitudes and
Flatt\'e parametrization for $f_0(980)$. See caption of Table
\ref{bwresult-1}. }
{\scriptsize\bc\renewcommand{\arraystretch}{1.3}\begin{tabular}{cccccc}
\hline\hline
&Sol.16\phantom{rrr}&Sol.17\phantom{rrr}&Sol.18\phantom{rrr}&Sol.19\phantom{rrr}&Sol.20\phantom{rrr}\\
\hline
\hspace{-2.5mm}$S_1 $\hspace{-4mm}&\hspace{-3.5mm} 0.73-i0.14\hspace{-3.5mm}&---\phantom{rrr}                         &---\phantom{rrr}                         &---\phantom{rrr}                         &---\phantom{rrr}                          \\
\hspace{-2.5mm}$S_2 $\hspace{-4mm}&\hspace{-3.5mm} 3.57+i5.18\hspace{-3.5mm}&\hspace{-3.5mm} 1.39+i4.12\hspace{-3.5mm}&\hspace{-3.5mm} 1.76+i4.10\hspace{-3.5mm}&\hspace{-3.5mm} 1.77+i3.27\hspace{-3.5mm}&\hspace{-3.5mm} 1.78+i3.76\hspace{-3.5mm} \\
\hspace{-2.5mm}$S_2'$\hspace{-4mm}&\hspace{-3.5mm} 2.53+i4.94\hspace{-3.5mm}&\hspace{-3.5mm}-1.33+i2.50\hspace{-3.5mm}&\hspace{-3.5mm}-1.17+i1.87\hspace{-3.5mm}&\hspace{-3.5mm}-2.60+i2.26\hspace{-3.5mm}&\hspace{-3.5mm}-2.97+i2.00\hspace{-3.5mm} \\
\hspace{-2.5mm}$S_3 $\hspace{-4mm}&\hspace{-3.5mm} 0.96-i0.58\hspace{-3.5mm}&\hspace{-3.5mm}-0.15+i0.67\hspace{-3.5mm}&---\phantom{rrr}                         &\hspace{-3.5mm}-2.03-i0.38\hspace{-3.5mm}&\hspace{-3.5mm}-0.81-i0.40\hspace{-3.5mm} \\
\hspace{-2.5mm}$S_4 $\hspace{-4mm}&\hspace{-3.5mm} 0.71-i0.71\hspace{-3.5mm}&\hspace{-3.5mm} 0.66-i0.61\hspace{-3.5mm}&\hspace{-3.5mm} 0.56-i0.70\hspace{-3.5mm}&---\phantom{rrr}                         &\hspace{-3.5mm} 0.29-i0.75\hspace{-3.5mm} \\
\hspace{-2.5mm}$S_5 $\hspace{-4mm}&\hspace{-3.5mm} 4.31-i1.36\hspace{-3.5mm}&\hspace{-3.5mm} 2.72-i0.92\hspace{-3.5mm}&\hspace{-3.5mm} 2.15-i1.35\hspace{-3.5mm}&\hspace{-3.5mm} 2.34+i1.82\hspace{-3.5mm}&---\phantom{rrr}                          \\
\hspace{-2.5mm}$V_2 $\hspace{-4mm}&\hspace{-3.5mm} 2.79+i1.89\hspace{-3.5mm}&\hspace{-3.5mm} 1.82+i1.83\hspace{-3.5mm}&\hspace{-3.5mm} 1.85+i1.82\hspace{-3.5mm}&\hspace{-3.5mm} 1.97+i1.89\hspace{-3.5mm}&\hspace{-3.5mm} 1.27+i2.15\hspace{-3.5mm} \\
\hspace{-2.5mm}$T_1 $\hspace{-4mm}&\hspace{-3.5mm}-5.39+i0.00\hspace{-3.5mm}&\hspace{-3.5mm}-5.55+i0.00\hspace{-3.5mm}&\hspace{-3.5mm}-5.55+i0.00\hspace{-3.5mm}&\hspace{-3.5mm}-5.80+i0.00\hspace{-3.5mm}&\hspace{-3.5mm}-5.61+i0.00\hspace{-3.5mm} \\
\hline
$\chi^2$                          &62.47\phantom{rrr}                       &65.53\phantom{rrr}                       &65.97\phantom{rrr}                       &73.20\phantom{rrr}                       &69.30\phantom{rrr}                        \\
\hline\hline \end{tabular}
\vspace{-5mm}
\renewcommand{\arraystretch}{1.0}\ec}\end{table}

Instead of using masses and widths from \cite{Anisovich:2002ij}, we may
chose values given in the PDG listings. The overall $\chi^2$
deteriorates by 2 units; the $\chi^2$ changes remain close to the ones
shown in Tables \ref{bwresult-1} and \ref{bwresult-2}. When the
$f_0(1370)$ parameters are changed, mass and width can adopt
nearly arbitrary values. Its parameters cannot be deduced from this
data. A free fit to $f_0(1710)$ properties puts its mass to the limit of
the phase space, with a marginal $\chi^2$ improvement.

The fractional contributions of the various isobars to the Dalitz
plot are not well defined quantities. First, interference effects
make it impossible to assign a fraction of a data to an {\it
amplitude}. The second argument is more technical: the fit fraction
depends on the model used. Tables \ref{kresult-pf},\ref{kresult-s}
and \ref{bwresult-1}, \ref{bwresult-2} show the spread of results
when the fit hypothesis is varied. From solution 14 we estimate that
nearly 70\% of the intensity comes from $f_0(980)$, 8\% from
$f_0(1500)$, 8\% from $\rho(1450)$, and 15\% from  $f_2(1270)$.

\subsection{Mass scans}

In mass scans, the mass of one resonance is changed in steps, and the
change of $\chi^2$ as a function of the imposed mass is inspected. In
a well behaved situation, the $\chi^2$ distribution exhibits a local
minimum at the optimal mass.

The resonance is produced with an arbitrary phase relative to the
production of another meson.  When the mass of the physical meson
under study and the Breit-Wigner test mass coincide (and their
respective widths), the Breit-Wigner amplitude and phase, and
amplitude and phase of the physical resonance agree over the full
mass range of the resonance. The overall phase difference of the
resonances under study relative to the other mesons is determined in
the fit; this phase is denoted here as $\varphi_0$. When the test
mass is changed, the Breit-Wigner amplitude and phase do no longer
match the complex amplitude in the data. In this situation, the
fitted amplitude adjusts its phase in a way that the true phase and
the fitted phase match in the mass range in which the true resonance
and the test Breit-Wigner amplitude have a large overlap. In a mass
scan, the phase motion is thus approximately reproduced by the
fitted phase. This method was developed in \cite{Klempt:2006sa}; a
first application showed that the $\eta(1440)$ region hosts one
un-split resonance \cite{Klempt:2004xg}. The analytical form of a
Breit-Wigner amplitude leads to an overall sign change of the
amplitude when going far below the nominal mass to far above the
resonance, independent of its coupling to other channels,
independent of inelasticities. The method explores directly the
phase of a Breit-Wigner amplitude.

\begin{figure}[pt]
\begin{center}
\begin{tabular}{cc}
\hspace{-1mm}\includegraphics[width=0.24\textwidth,height=0.22\textwidth]{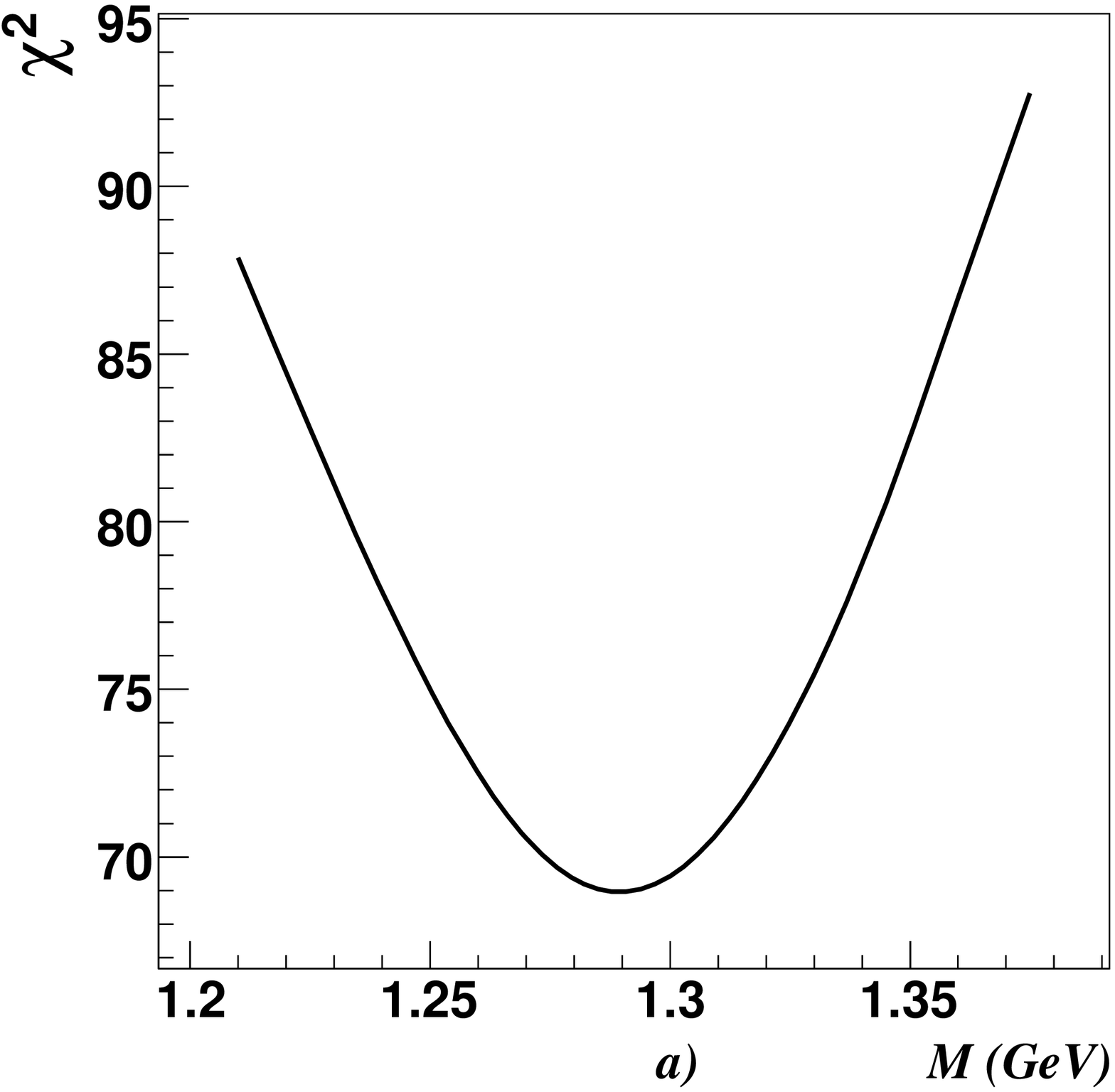}&
\hspace{-5mm}\includegraphics[width=0.24\textwidth,height=0.22\textwidth]{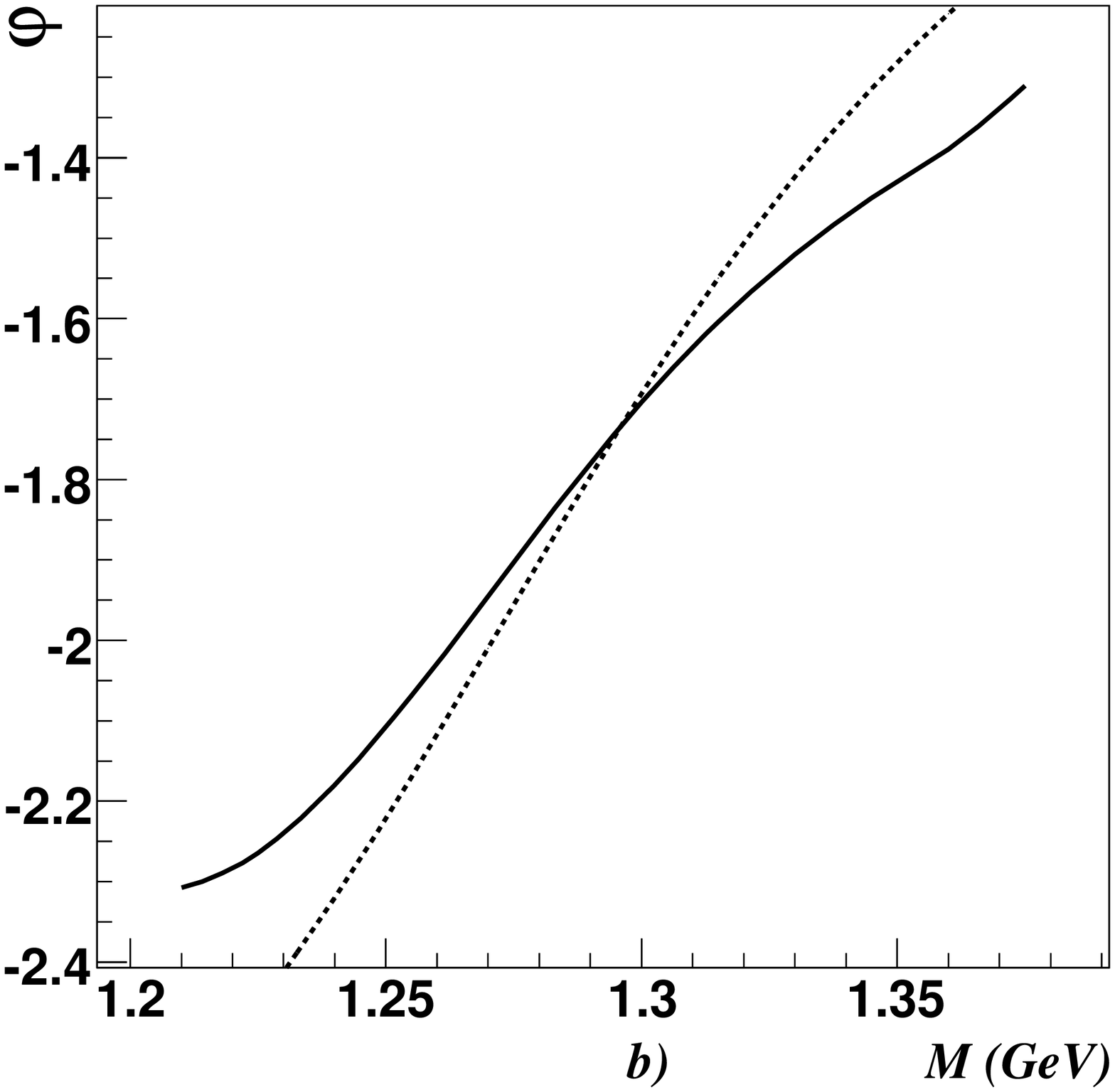}
\end{tabular}
\vspace{-2mm}
\end{center}
\caption{\label{scan-f2}
a: $\chi^2$ of a fit in which the mass of the $f_2(1270)$ is scanned.
b: The associated phase as a function of the mass. }
\end{figure}

We test this idea by monitoring the phase of the $f_2(1270)$. Fig.
\ref{scan-f2} shows a scan of the $f_2$ mass region. The minimum in
$\chi^2$ reached at 1280\,MeV (see Fig. \ref{scan-f2}b). The 'observed'
phase resulting from the fit shows a strong variation as a function of
the imposed $f_2$ mass, stronger than expected for a single
Breit-Wigner. A phase variation is observed as expected but
quantitatively, the expected phase variation is not reproduced. This
may serve as a warning that the method is not without risk.

\begin{figure}[pb]
\begin{center}
\begin{tabular}{cc}
\hspace{-1mm}\includegraphics[width=0.24\textwidth,height=0.22\textwidth]{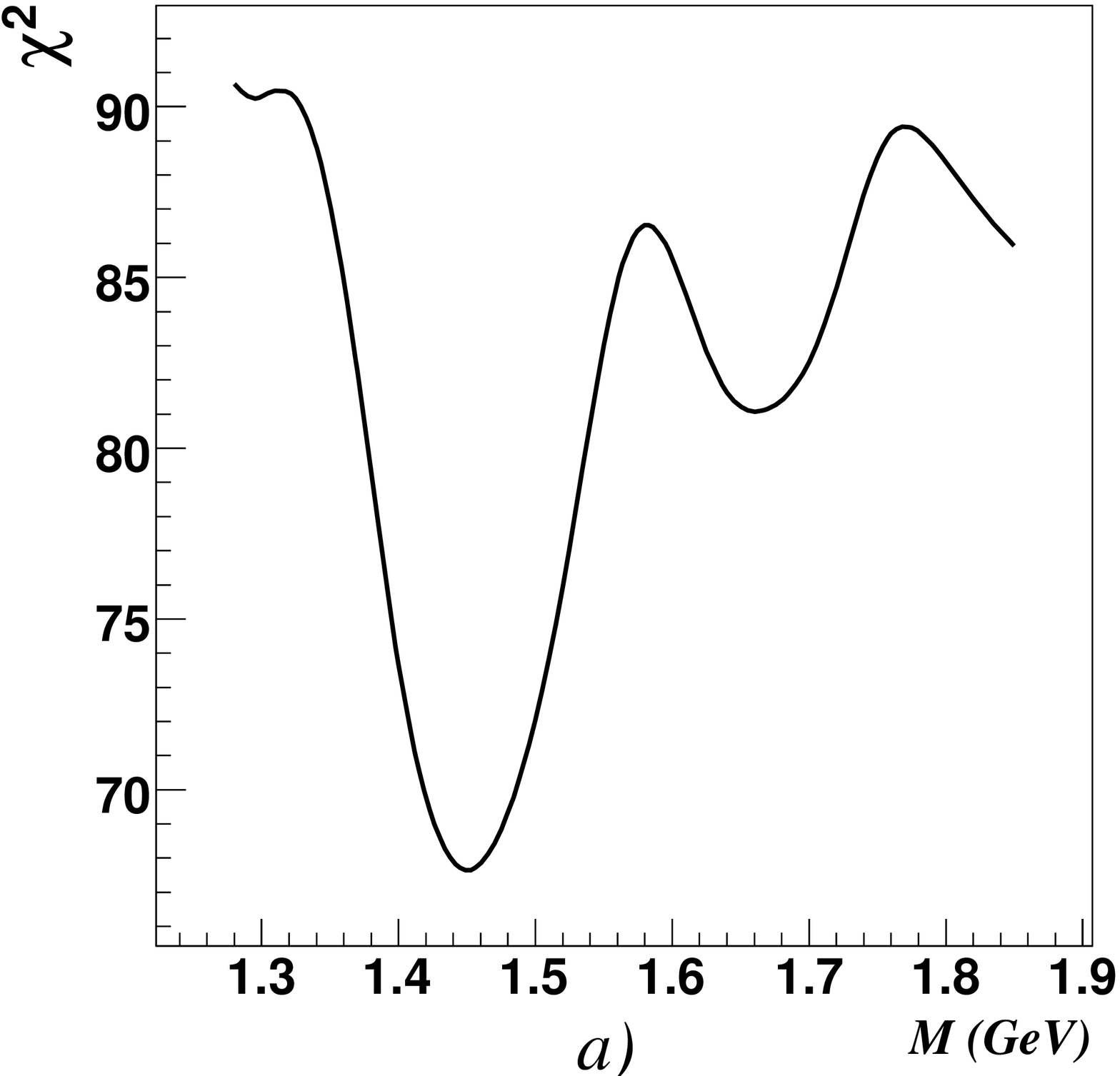}&
\hspace{-5mm}\includegraphics[width=0.24\textwidth,height=0.22\textwidth]{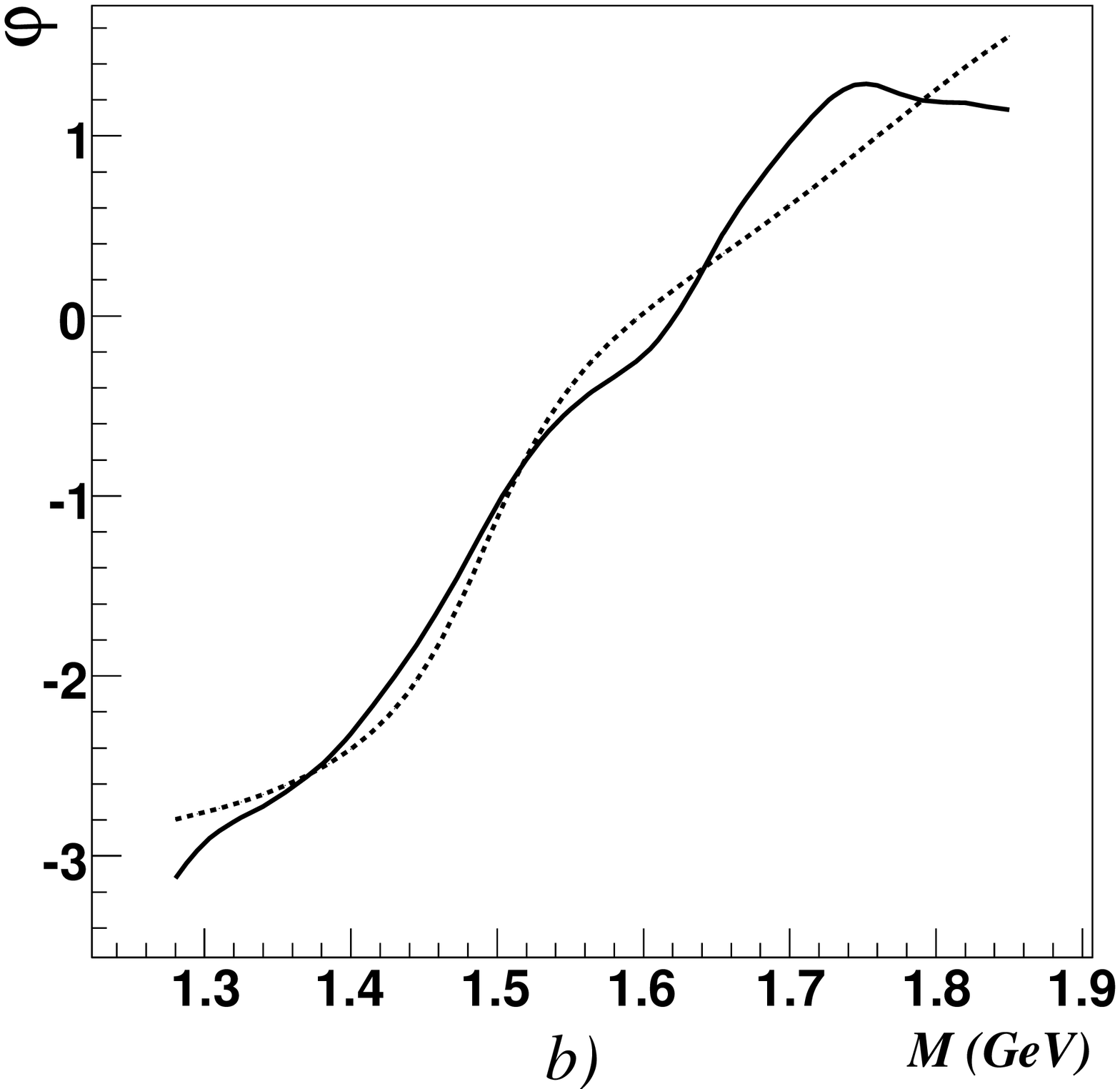}
\end{tabular}
\vspace{-2mm}
\end{center}
\caption{\label{scan-one}
a: $\chi^2$ of a fit which includes $f_0(980)$, $\rho(1450)$, and
$f_2(1270)$ with fixed masses and widths and one scalar resonance the
mass of which is scanned. The width of this resonance is fixed to
110\,MeV. b: The scalar phase motion as derived from the fit (solid
line). The fitted scalar phase motion (dashed line) assumes two
additional scalar resonances, $f_0(1500)$ and $f_0(1710)$. }
\end{figure}

Fig. \ref{scan-one}a shows the scan of a scalar amplitude.
We start from solution 13 of Table \ref{bwresult-1}, fix the mass
of the scalar meson at pre-set values, and determine the $\chi^2$ as a
function of the pre-set mass.  The scalar mass is varied from 1.2 to
1.7\,GeV/c$^2$. A deep minimum in $\chi^2$ is observed at $M=1452\pm
22$\,MeV/c$^2$ reproducing findings of previous analyses. At a mass of
1680\,MeV, a weaker but still pronounced local minimum is seen which we
tentatively identify with $f_0(1710)$. The mass shift with respect to
the expected value is possibly due to the phase space limitation.

The phase motion observed in the scan and shown in Fig. \ref{scan-one}b
exhibits a surprise: in the range from 1.2 to 1.7\,GeV/c$^2$ it covers
more than $2\pi$. The region must house more
than a single resonance.
This result encouraged us to introduce a $f_0(1710)$ with fixed
parameters. The phase motion expected for two resonances (plus a
small contribution from $f_0(980)$) agrees reasonably well with
the 'measured' phase, see Fig. \ref{scan-one}b.

\begin{figure}[pt]
\begin{center}
\begin{tabular}{cc}
\hspace{-5mm}\includegraphics[width=0.24\textwidth,height=0.22\textwidth]{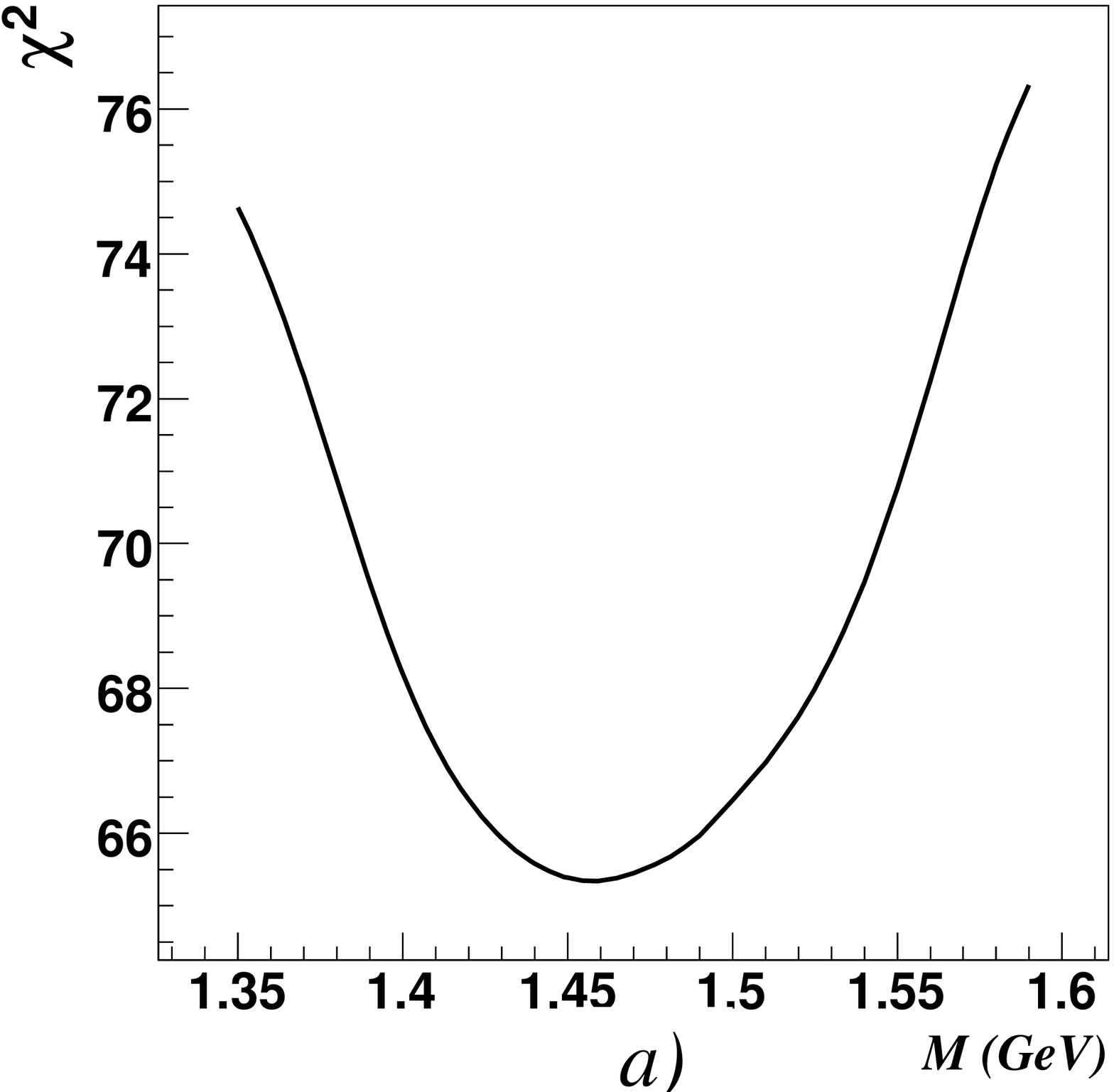}&
\hspace{-5mm}\includegraphics[width=0.24\textwidth,height=0.22\textwidth]{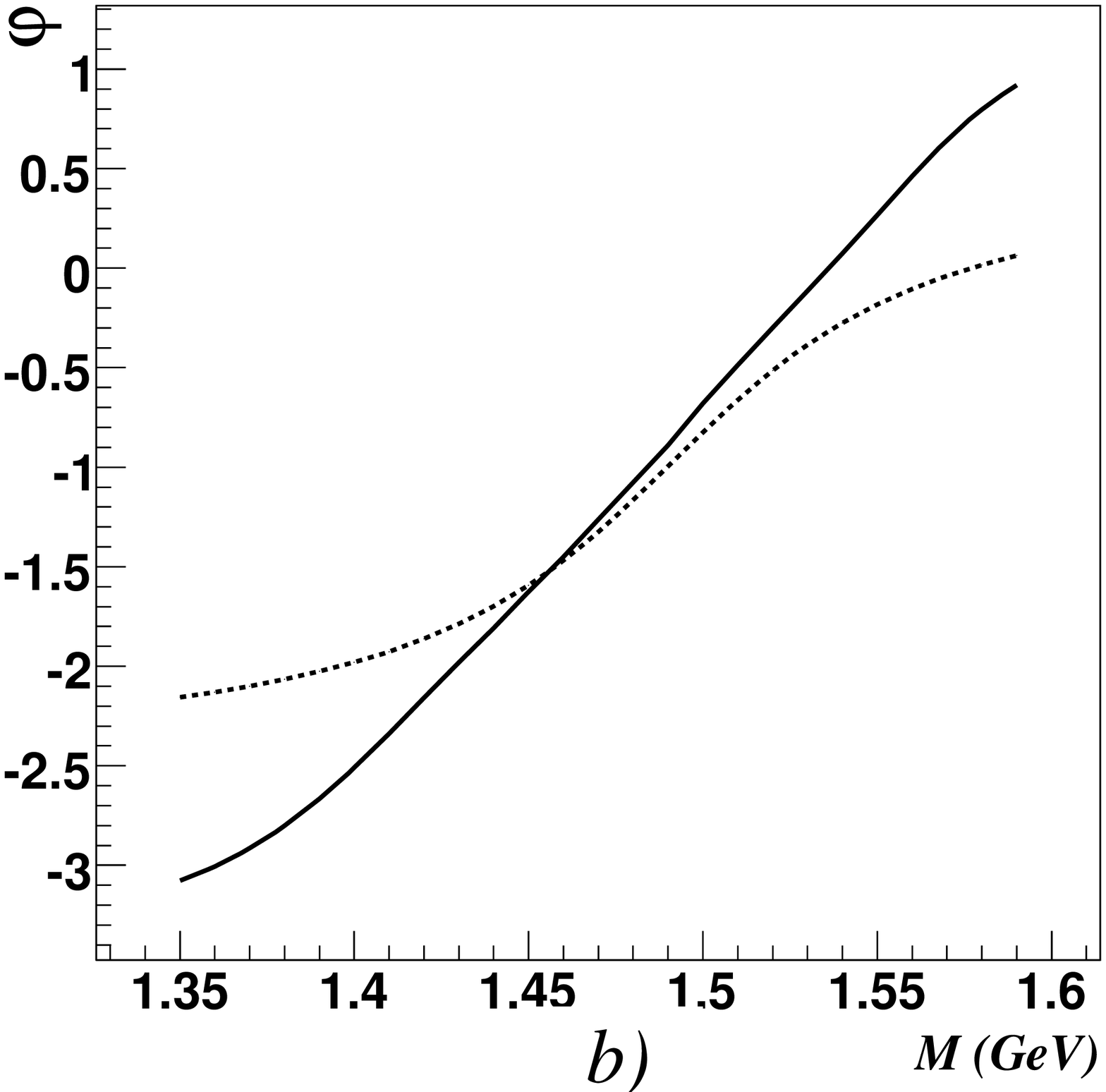}
\end{tabular}
\vspace{-2mm}
\end{center}
\caption{\label{scan-two} a: $\chi^2$ of a fit which includes
$f_0(980)$, $\rho(1450)$, $f_2(1270)$, and  $f_0(1710)$ with fixed
masses and widths and one scalar resonance the mass of which is
scanned. The width of this resonance is fixed to 110\,MeV. b: The
scalar phase motion as derived from the fit (solid line). The fitted
scalar phase motion (dashed line) assumes one additional scalar
resonance, $f_0(1500)$. The `observed' phase motion covers a wider
range than a single resonance does.}
\end{figure}
As a next step, we include $f_0(1710)$ in the fit with fixed
parameters and study the remaining scalar wave in a scan with a
second scalar resonance. The result is shown in Fig.
\ref{scan-two}a, the associated phase motion in  Fig.
\ref{scan-two}b. There is now one $\chi^2$ minimum which is wider
than the central minimum seen in Fig. \ref{scan-one}a.  When
$f_0(1710)$ is added, the mass uncertainty increases, and the fit
yields $M=1470\pm 60$\,MeV/c$^2$, a value which is not incompatible
with 1500\,MeV. If instead of $f_0(1710)$ the $f_0(1370)$ resonance
is included, the production strength of the second scalar resonance
gets small, and the $\chi^2$ minimum is found at 1400\,MeV.
Obviously, the interference of two close-by resonances can mimic
the observed structure but we discard this solution.

The phase motions shown in Figs. \ref{scan-one}b and \ref{scan-two}b
are suggestive but should be discussed with some reservations. The
observed phase is certainly influenced by the data but the fit does
not determine the local phase. The phase is derived from a
comparison of data and a Breit-Wigner amplitude of finite (110\,MeV)
width; thus, the phase motion is smeared out. Second, the phase is
determined from fits in which the mass of the resonance under
studied is intentionally detuned. It is hence not guaranteed that
the fit does not explore different local minima. In order to test
the method, we have made the scans with different widths of the test
Breit-Wigner amplitude (with $\Gamma=60$ or $160$\,MeV/ c$^2$). No
significant changes were found supporting the hypothesis that the
observed phase motions reflect some physics content.

\begin{figure}[pt]
\begin{center}
\begin{tabular}{cc}
\hspace{-5mm}\includegraphics[width=0.24\textwidth,height=0.22\textwidth]{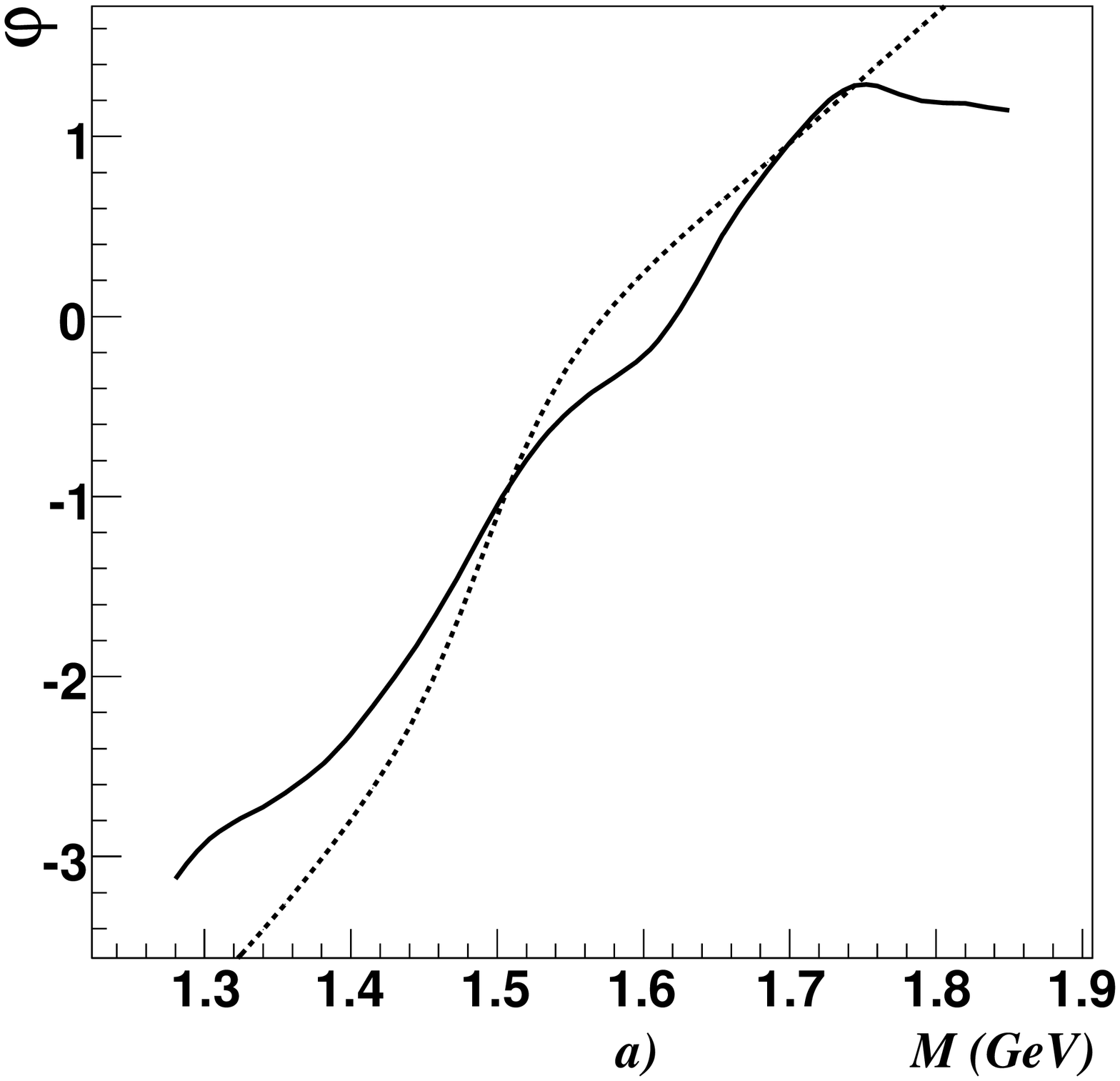}&
\hspace{-5mm}\includegraphics[width=0.24\textwidth,height=0.22\textwidth]{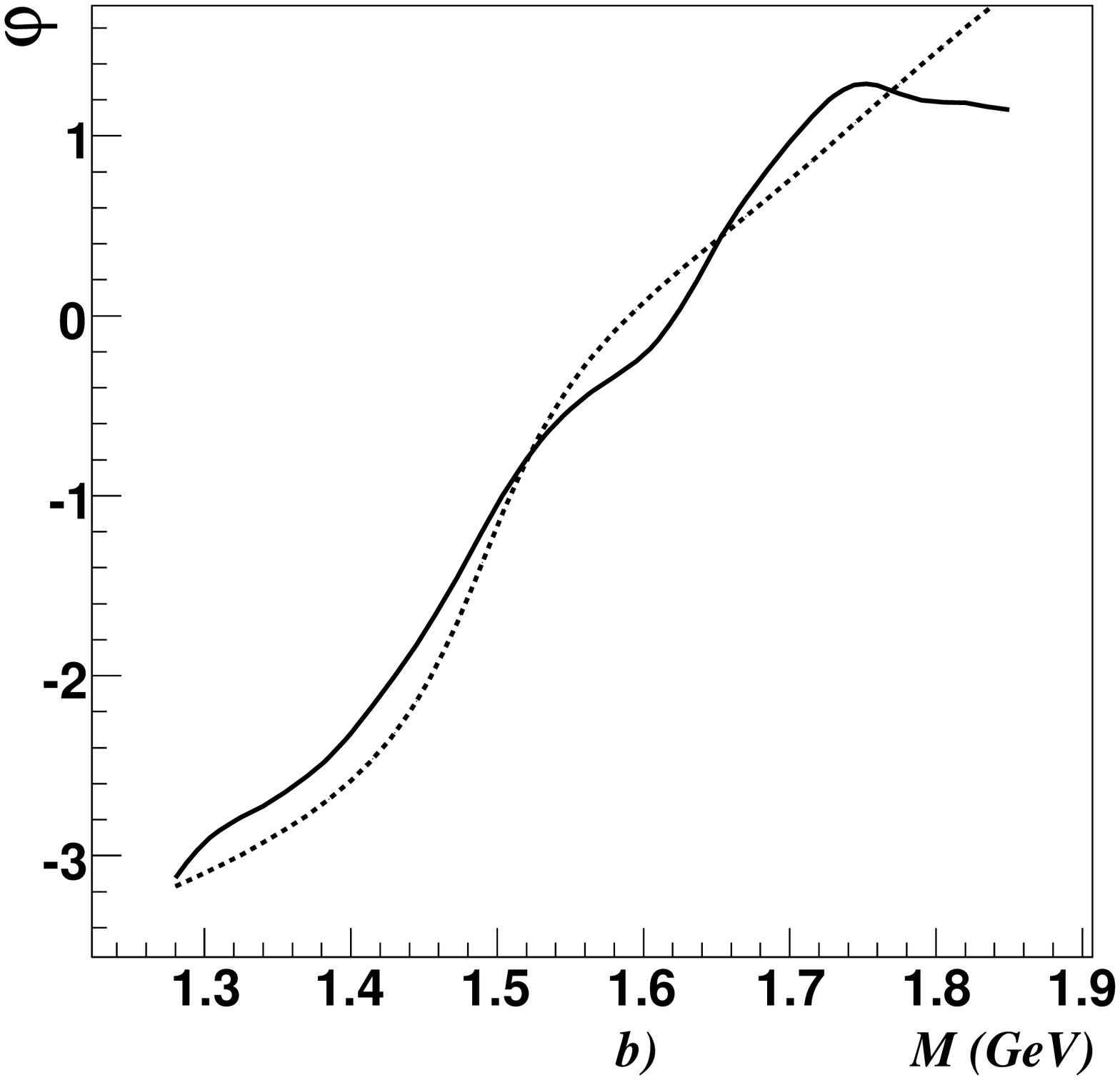}
\end{tabular}
\vspace{-2mm}
\end{center}
\caption{\label{scan-phase}
The scalar phase motion as derived from the fit (solid
line). The fitted scalar phase motion (dashed line) assumes three
additional scalar resonances. In (a), these are $f_0(1370)$,
$f_0(1500)$, and $f_0(1710)$; in (b) $f_0(1000)$, $f_0(1500)$, and
$f_0(1710)$, where  $f_0(1000)$ represents the wide scalar background
discussed in the text.} \end{figure}

Thus we go one step beyond, and try to determine the number of
resonances by comparing the observed phase motions with Breit-Wigner
phase motions of resonances with fixed masses and widths. In Fig.
\ref{scan-phase}, the `observed' phase motion of Fig. \ref{scan-one}
is shown together with two fits. Both fits use $f_0(1500)$ and
$f_0(1710)$, fit (a) adds $f_0(1370)$, fit (b) adds the wide
background amplitude  $f_0(1000)$ suggested by Au, Morgan, and
Pennington \cite{Au:1986vs}, and by Min\-kowski and Ochs
\cite{Minkowski:1998mf}. Both possibilities reproduce the `observed'
phase shift reasonably well even though $f_0(1370)$ leads to an
expected phase motion exceeding slightly the observed one while
introducing $f_0(1000)$ leads to a perfect match.

We have tried to perform a mass-independent analysis by choosing complex
amplitudes for the scalar partial wave in each mass bin, and offering
these to the minimization process. This mass independent partial wave
amplitude was partly restricted to 2, 3, or 4 mass bins. The results
were unstable; we failed to derive a mass independent partial wave
amplitude from these studies.

\section{Discussion and conclusions}

We have studied the Dalitz plot for $D^+_s$ decays into $\pi^- \pi^+
\pi^+$. In agreement with earlier findings, we found that the
largest fraction of the data stems from scalar isoscalar mesons
decaying into $\pi^-\pi^+$, in particular the $f_0(980)$ meson. The
$\rho(1460)$ and $f_2(1270)$ contribute significantly as well. We
have made the attempt to understand the conflicting results on
scalar mesons which were obtained from fits to the
$D_s^+\to\pi^-\pi^+\pi^+$ Dalitz plot. Previous fits agree that the
data require scalar intensity in the mass region above the
$f_0(980)$. When fitted with a Breit-Wigner amplitude, an optimal
mass between 1440 and 1475\,MeV/c$^2$ was found. The masses quoted
were not compatible with the mass of neither the $f_0(1370)$ --
which we rather assume to have 1320\,MeV/c$^2$ -- nor with the
$f_0(1500)$ resonance. Thus it is often argued that both,
$f_0(1370)$ and $f_0(1500)$, might be required to get a good fit
when nominal (PDG) masses and widths are imposed.

We reproduce these results. However, when a high mass resonance,
$f_0(1710)$, is introduced, the $f_0(1500)$ resonance alone is
sufficient to yield an acceptable fit. The $\chi^2$ as a function of
the assumed scalar mass of the $f_0(1370)/f_0(1500)$ develops a flat
floor, and the $1\sigma$ mass interval extends from 1.41 to
1.53\,MeV/c$^2$. Thus the observed mass is fully compatible with the
hypothesis that the standard $f_0(1500)$ is produced in the
reaction. An additional contribution from a $f_0(1370)$ resonance is
not required. If the existence of $f_0(1370)$ is assumed, its
parameters can be chosen to agree with the `narrow' $f_0(1370)$ of
the PDG or with the `wide' red dragon of Minkowsky and Ochs. The
statistics we used is not sufficient to discriminate between the
three alternatives `no $f_0(1370)$', `wide $f_0(1370)$', or `narrow
$f_0(1370)$'. The $\sigma(500)$ leads to a marginal improvement of
the fit, with a statistical evidence just above one standard
deviation. In the $K$-matrix fits, the inclusion of the $f_0(1370)$
resonance was necessary in fits to a large body of different
reactions. In the fits to the $D_s\to 3\pi$ data, $f_0(1370)$ does
provide a notable improvement even though this data alone is not
sufficient to claim its existence.

One could dream of future high-statistics high-quality data. There
is however the possibility to combine existing data from different
experiments. The reaction $D_s^+\to \pi^- \pi^+ \pi^+$ was studied
at Fermilab by the E687 (434 events) \cite{Frabetti:1997sx}, E791
(848 events) \cite{Aitala:2000xt} and the FOCUS (1475 events)
\cite{Link:2003gb,Malvezzi:2004tq} collaborations. The BaBar
collaboration has extracted 2900 events; preliminary results were
reported in a PhD thesis at Bochum \cite{Deppermann}. Including
Belle results, a factor 10 can be reached in statistics. The
different background contributions would help to understand the
systematic errors. We urge that enterprize should be undertaken.

\section*{Acknowledgements}
We thank the E791 collaboration for providing us with their data
described in \cite{Aitala:2000xu}, for simulations of their detector
(for acceptance corrections) and for other related discussions. Helpful
comments by B. Meadows and A. Correa dos Reis are particularly
acknowledged. M.M. was supported by a grant from the Deutsche
Forschungsgemeinschaft.

 \small \bibliographystyle{unsrt}

\end{document}